# The MOBH35 Metal–Organic Barrier Heights Reconsidered: Performance of Local-Orbital Coupled Cluster Approaches in Different Static Correlation Regimes

Emmanouil Semidalas and Jan M.L. Martin*

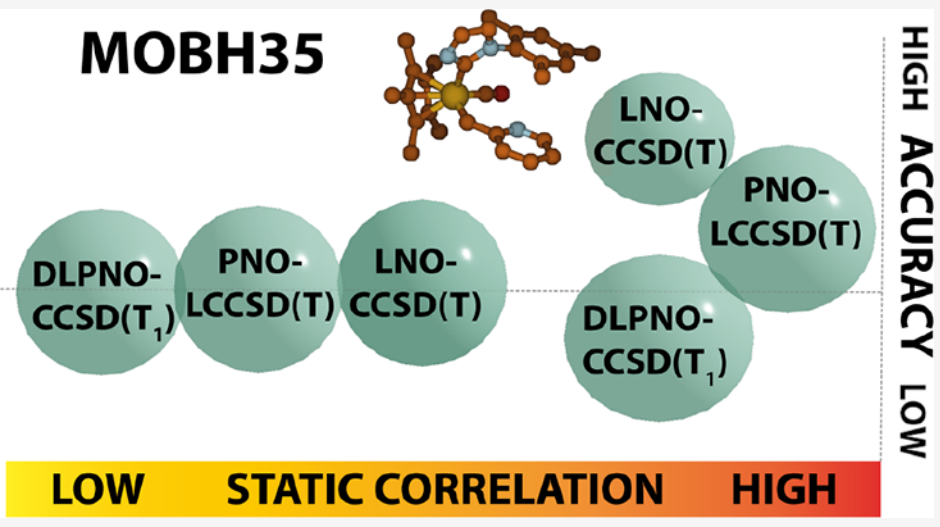

**ABSTRACT:** We have revisited the MOBH35 (Metal–Organic Barrier Heights, 35 reactions) benchmark [Iron, Janes, *J. Phys. Chem. A*, **2019**, *123* (17), 3761–3781; ibid. **2019**, *123*, 6379–6380] for realistic organometallic catalytic reactions, using both canonical CCSD(T) and localized orbital approximations to it. For low levels of static correlation, all of DLPNO-CCSD(T), PNO-LCCSD(T), and LNO-CCSD(T) perform well; for moderately strong levels of static correlation, DLPNO-CCSD(T) and ($T_1$) may break down catastrophically, and PNO-LCCSD(T) is vulnerable as well. In contrast, LNO-CCSD(T) converges smoothly to the canonical CCSD(T) answer with increasingly tight convergence settings. The only two reactions for which our revised MOBH35 reference values differ substantially from the original ones are reaction 9 and to a lesser extent 8, both involving iron. For the purpose of evaluating density functional theory (DFT) methods for MOBH35, it would be best to remove reaction 9 entirely as its severe level of static correlation makes it just too demanding for a test. The magnitude of the difference between DLPNO-CCSD(T) and DLPNO-CCSD($T_1$) is a reasonably good predictor for errors in DLPNO-CCSD($T_1$) compared to canonical CCSD(T); otherwise, monitoring all of $T_1$, $D_1$, max$|t_i^A|$, and $1/(\varepsilon_{\text{LUMO}} - \varepsilon_{\text{HOMO}})$ should provide adequate warning for potential problems. Our conclusions are not specific to the def2-SVP basis set but are largely conserved for the larger def2-TZVPP, as they are for the smaller def2-SV(P): the latter may be an economical choice for calibrating against canonical CCSD(T). Finally, diagnostics for static correlation are statistically clustered into groups corresponding to (1) importance of single excitations in the wavefunction; (2a) the small band gap, weakly separated from (2b) correlation entropy; and (3) thermochemical importance of correlation energy, as well as the slope of the DFT reaction energy with respect to the percentage of HF exchange. Finally, a variable reduction analysis reveals that much information on the multireference character is provided by $T_1$, $I_{\text{ND}}/I_{\text{tot}}$, and the exchange-based diagnostic $A_{100}$[TPSS].

## ■ INTRODUCTION

Reactions of transition metal (TM) complexes are of the greatest importance for progress in sustainable catalysis,[1] energy storage,[2] biological chemistry,[3] drug development,[4] and many other research areas. For their computational modeling and design (see a recent review),[5] density functional methods are widely used, but wavefunction ab initio methods are increasingly used for calibration of DFT approaches.

The single-reference CCSD(T) [coupled cluster with all singles and doubles, augmented with quasiperturbative triple excitations[6,7]] has a well-deserved reputation as a "gold standard" for systems where electron correlation is predominantly dynamic, i.e., where the Hartree–Fock solution is a reasonable zero-order approximation. To a large extent, it benefits from error cancellation[8−12] between neglect of (usually antibonding) higher-order connected triple excitations (as accounted for in full CCSDT[13]), which tend to be repulsive, and complete neglect of (universally bonding) connected quadruple excitations, as treated in the CCSDT(Q)[14] or full CCSDTQ[15,16] methods. (For an alternate perspective from perturbation theory, see Stanton's work.[17])

CCSD(T) is comparatively "black-box" (requiring zero or minimal operator "judgment calls") and has "only" $n^3N^4$ asymptotic CPU time scaling (with $n$ the number of electrons and $N$ the size of the basis set), compared to $n^3N^5$ for CCSDT and $n^4N^6$ for CCSDTQ. Additionally, however, a number of localized orbital approximations to CCSD(T) have been developed[18−23] (see below) that in principle scale linearly with system size and hence could be applied directly to the catalysts of interest. For main-group elements, we recently showed[24,25] that when local wavefunction methods are embedded into composite wavefunction protocols, their accuracy is comparable to the reference energies in the GMTKN55 database[26] of 1500 reactions for species of main-group elements.

There has been a number of recent benchmark studies using coupled cluster methods on dissociation energies of transition metal systems, such as a recent update and extension[27] of the 20 metal−ligand bond energies dataset,[28] 3dMLBE20, or the copper, silver, and gold clusters' benchmarks.[29,30] It has been argued[28] that any differences between experimental and CCSD(T) dissociation energies of diatomic 3d TMs lie in the expected range once the experimental energies are revised.[31]

However, in catalysis, one is more interested in complexation energies and barrier heights of "real-life" organometallic systems. To this end, Dohm et al.[32] developed the MOR41 (41 metal−organic reactions) benchmark, and Iron and Janes[33,34] the MOBH35 benchmark of forward and reverse barrier heights for 35 representative metal−organic reactions. In both cases, reference data were obtained by complete basis set extrapolation of DLPNO-CCSD(T) data (domain localized pair natural orbital coupled cluster[18,19,35,36]); Iron and Janes additionally considered[34] the improved DLPNO-CCSD($T_1$) approach. Both sets of authors then proceeded to assess the performance of a large number of density functional and approximate wavefunction approaches. Most recently, the work of Dohm et al. was followed up by a similar study[37] on open-shell transition metal reactions, where DLPNO-CCSD($T_1$)/CBS (complete basis set extrapolated) benchmarks were presented and used for the assessment of many DFT and semiempirical methods.

In a recent study on polypyrroles[38] (extended porphyrins where Hückel and bicyclic structures are mostly "single-reference", but Möbius structures have pronounced static correlation), we, however, found that the performance of different localized approaches depends strongly on the degree of static correlation in the system. In particular, we found that the local natural orbital coupled cluster, LNO-CCSD(T),[21,39−42] proved to be considerably more resilient to static correlation than the DLPNO-CCSD($T_1$) and PNO-LCCSD(T)[20,43,44] approaches. At least a subset of the MOBH35 reactions can be expected to exhibit significant static correlation.

In the present work, we will carry out canonical CCSD(T) calculations for a large subset of MOBH35, evaluate the performance of different localized methods, and attempt to rationalize this by means of a large number of diagnostics for static correlation. As a byproduct, updated reference data for MOBH35 are obtained. It is shown that, by and large, the DLPNO-CCSD($T_1$)-based reference data are adequate for evaluating all but the most accurate DFT functionals. However, for one particular system (the product of reaction **9**), static correlation is severe enough that it causes a catastrophic breakdown of localized approaches. To be fair, that case is arguably well beyond the safe use margin of CCSD(T) itself.

## ■ COMPUTATIONAL METHODS

Geometries for reactants, transition states, and products of reactions **1−9**, **11−16**, and **21−34** were taken verbatim from the Supporting Information (SI) of Dohm et al.,[45] except that the product geometry of reaction **5** (which erroneously is the same as the reactant geometry in the said SI) was taken instead from Iron and Janes,[33] as were the structures for the bimolecular reactions **10** and **35**. The bimolecular reactions **17−20** involved transition states too large for canonical calculations and were hence omitted.

The Weigend−Ahlrichs basis set family[46] of def2-SV(P), def2-SVP, def2-TZVP, def2-TZVPP, and def2-QZVPP was used throughout, as it was designed as a compromise between the requirements of wavefunction and DFT calculations. (We note that for reactions **1−9**, which involve first-row transition metals (TMs), they do not contain ECPs; for the second- and third-row TMs, they contain Stuttgart RECPs.[47]) Where relevant, the standard def2-JK fitting basis set[48] was used for Coulomb and exchange integrals, and the standard def2-*n*ZVPP-RI[49,50] basis sets for RI-MP2 and RI-CCSD calculations.

Where possible, conventional canonical CCSD(T)[6,7] calculations with the def2-SVP basis set were carried out using MOLPRO 2020 and 2021,[51] devoid of any fitting basis sets. For the largest species, involved in reactions **1**, **2**, **5**, **8**, and **9**, RI-CCSD(T) was instead carried out using the algorithm[52] in PSI4 version 1.4.[53] For verification purposes, we also reran the smaller species with this algorithm and compared them with the conventional results as well as with a newer algorithm[54] in MRCC 2020.[55] (For reactions **3−4** and **6−7**, energies obtained with PSI4 and MRCC perfectly match, while differences with the canonical answers were negligible. For some second- and third-row TMs, there were subtle differences between PSI4 and MOLPRO that eventually could be narrowed down to the SCF reference energy and hence to subtle differences in the ECP parameters; no such issue was found for MRCC.) Convergence to the lowest closed-shell singlet SCF solution was verified by comparing the SCF energies from each program with those obtained using Gaussian 16[56] in the same basis set. SCF thresholds of $10^{-9}$ $E_h$ were used in MOLPRO and ORCA, while the default thresholds of $10^{-6}$ $E_h$ in the SCF energy and $10^{-7}$ for the RMS change in the density matrix were employed in MRCC; in practice, the energies are converged 2−3 decimal places better than the energy threshold because of the stringent density criterion.

Three different families of localized orbital approaches were considered:

(a) MOLPRO was used for the PNO-LCCSD(T) localized orbital coupled cluster calculations,[20,44,57,58] both using built-in "default" and "tight" cutoff parameter ensembles (see Table 1 in ref 20 for details).

(b) For LNO-CCSD(T),[41,42,59] the implementation in MRCC 2020[55] was employed using the Normal, Tight, and vTight cutoff parameter ensembles, as detailed in Table 1 of ref 59. In addition, we considered "Tight+", which combines the vTight value of $10^{-6}$ $E_h$ for wpairtol with Tight values for the other parameters: in a recent study on polypyrrole isomerization,[38] we found this to perform almost as well as vTight.

(c) DLPNO-CCSD(T)[60] and the version with improved iterative triples, DLPNO-CCSD($T_1$),[36] were calculated using ORCA[61] 4.1 and 4.2, using NormalPNO and TightPNO settings (see Table 1 in ref 35 for details). We also considered VeryTightPNO settings as proposed by Pavošević et al.:[62] TCutPNO = $10^{-8}$, TCutMKN = $10^{-4}$, and TCutPairs = $10^{-6}$ but leaving TCutDo at its automatically determined value as recommended by the ORCA manual and applied in ref 37 [A. Hansen, personal communication]. In all these calculations, the RIJCOSX chain-of-spheres approximation[63,64] was applied in conjunction with ORCA4's tightest COSX grid (GRIDX9).

Unless stated otherwise, ORCA "chemical cores" were frozen throughout, i.e., valence electrons only were correlated except for the $(n-1)$sp electrons on the metal. There are 10 inner shell electrons for the 3d elements, and the associated ECPs were employed for the 4d and 5d elements as mentioned above (see

Table S3 in the SI for the specific numbers of frozen core electrons for each species in this study).

Complete basis set extrapolation (CBS) parameters for the Weigend−Ahlrichs basis sets were taken from Table 3 in Neese and Valeev.[65] For CCSD(T), this is the well-known partial-wave-like formula[66−70] $E(L) = E_{CBS} + L^{-\alpha}$, where $\alpha_{2,3} = 2.40$ for the $L = \{2,3\}$ pair def2-SVP and def2-TZVPP and $\alpha_{3,4} = 2.97$ for the $L = \{3,4\}$ pair def2-TZVPP and def2-QZVPP (functionally equivalent to $\alpha = 3$ in the simple Helgaker formula[71]). For the Hartree−Fock component, they used $E(L) = E_{CBS} + \exp(-\beta\sqrt{L})$, which they attributed to Karton and Martin[72] but actually was first proposed by Klopper and Kutzelnigg;[73−75] for the def2-{T,Q}ZVPP basis set pair, their Table 3 lists $\beta = 7.88$. Any two-point complete basis set extrapolation can be rewritten as $E_{CBS} = E[L] + A_{L-1,L,\text{method}}(E[L] - E[L-1])$ where the Schwenke coefficient[76] $A_{L-1,L,\text{method}}$ is specific to the basis set pair and the method. The above extrapolations correspond to $A_{2,3,\text{CCSD(T)}} = 0.607$, $A_{3,4,\text{CCSD(T)}} = 0.741$, and $A_{3,4,\text{HF}} = 0.1377$. (The latter would correspond to $\alpha = 7.34$. See ref 77 for a discussion of the inter-relations and equivalences between different two-point extrapolation formulas.)

A number of diagnostics for nondynamical correlation have been considered. One group relates to the single excitation amplitudes. First of all, there is the well-known $T_1$ diagnostic of Lee and Taylor,[78] which is defined as the Euclidean vector norm (2-norm) of the CCSD single amplitudes vector, divided by the square root of the number of correlated electrons, as follows:

$$T_1 = \frac{||\boldsymbol{t}_1||}{\sqrt{N}} = \sqrt{\frac{\boldsymbol{t}_1^{\mathrm{T}}\boldsymbol{t}_1}{N}}$$

(The denominator ensures that $T_1$ is size-intensive.) The $D_1$ diagnostic[79−81] instead is based on a matrix norm, namely, $(\max \lambda[\mathbf{t}_1 \cdot \mathbf{t}_1^{\mathrm{T}}])^{1/2}$ the square root of the largest eigenvalue of the outer product. It can easily be seen from an example like BN...*n*-octadecane that the large $T_1$ diagnostic of singlet boron nitride will be "quenched" by the aliphatic chain, while $D_1$ will be that of the worst fragment. (The ratio $D_1/T_1$, or rather its deviation from $2^{1/2}$, has been proposed as a measure for homogeneity of the system.[81]) Finally, $\max|t_1|$ or the largest single excitation amplitude in absolute value can also be seen as $|\mathbf{t}_1|_\infty$ the infinity-norm of $\mathbf{t}_1$.

From the double excitation amplitudes, the $D_2$ diagnostic[79] has been defined analogously to $D_1$, and of course, we have $\max|t_2| = |\mathbf{t}_2|_\infty$.

The pitfalls of relying on just a single diagnostic have been discussed at length in ref 82. It is sufficient to say here, for example, that the $F_2$ molecule has deceptively low $T_1 = 0.011$ and $\max|T_1| = 0.02$ but large $D_2 = 0.23$ and $\max|T_2| = 0.17$ on account of a prominent double excitation.

$-T_e$(SCF) is minus the first excitation energy from a STABLE calculation using Gaussian[56] 16. This number will be positive for a closed-shell molecule that has an RHF-UHF instability (which is itself an indicator of type A static correlation[83] a.k.a. absolute near-degeneracy correlation, such as in a molecule stretched past its Coulson−Fischer point[84] a.k.a. the RHF-UHF bifurcation point). Alternatively, it is straightforward to perform the stability analysis in PSI4 or considering CIS or TD-HF that are more or less equivalent. For RHF-UHF instabilities, the same result can be obtained, slightly less conveniently, by separate RHF and UHF with a HOMO−LUMO mixed guess (or Fermi smearing guess[85,86]).

In the "W4 theory" paper,[87] a number of energetics-based diagnostics were proposed. One was %TAE[(T)], the percentage of the total atomization energy accounted for by the connected triple excitations: it was found repeatedly[87−89] that this is a very good indicator for the importance of post-CCSD(T) contributions. The other was %TAE[SCF], the percentage of the atomization energy recovered at the Hartree−Fock level, which is about 70−80% for molecules dominated by dynamical correlation but drops as static correlation becomes more important and actually becomes negative (molecule metastable in the absence of correlation) for such cases as $F_2$ and $O_3$. We could of course instead look at its converse, $\%TAE_{corr} = 100\% - \%TAE[SCF]$, which will increase with increasing levels of static correlation.

A number of diagnostics are derived from natural orbital occupation numbers $n_i$, such as the Truhlar $M$ diagnostic,[90] which for closed-shell molecules reduces to simply $(2 - n_{HOMO} + n_{LUMO})/2$, Matito's nondynamical correlation index[91,92] $I_{ND}$, and the size-intensive variation that we proposed,[93] $r_{ND} = I_{ND}/I_{tot} = I_{ND}/(I_{ND} + I_D)$. Very recently, Matito et al. proposed[94] another size-intensive variation, $I_{ND}/N_{val}^{1/2}$.

In terms of DFT-based diagnostics, we considered the fractional orbital density (FOD), obtained from the fractional orbital occupations generated by a Fermi smearing calculation at a high finite temperatures.[95] (They were evaluated using ORCA's built-in procedure with default settings.)

Additionally, we considered the $A$ diagnostic proposed by Fogueri et al.,[82] which is effectively the slope of the DFT total atomization energy with respect to the percentage of Hartree−Fock exchange. We also evaluated the recently proposed %TAE[ΔX,TPSS] diagnostic,[96] which measures the difference in the exchange energy when substituting the HF density. (This was inspired by the classic work of Handy and Cohen,[97] who argued that the DFT exchange energy captures static correlation effects that the HF exchange energy intrinsically cannot.) In the present case, it is defined as %TAEx[TPSS@HF − HF] = 100%(TAE[X,TPSSx] − TAE[X,HF])/TAE[CCSD(T)].

Finally, we should mention diagnostics based on the coefficient of the HF determinant (in a method with regular normalization, like CASSCF or CI) or the sum of squares of all the excitation amplitudes (in intermediate normalization, like in coupled cluster theory). However, as pointed out by a reviewer to ref 38, these quantities are not size-intensive: as a remedy, we used here $2\ln(A)/N$, which remains constant for an assembly of any number of identical systems at infinite separation.

## ■ RESULTS AND DISCUSSION

Our best estimated forward and reverse barrier heights were obtained by a three-layer extrapolation process:

(a) The CCSD component was obtained by extrapolation to the infinite basis set limit from def2-TZVPP and def2-QZVPP basis sets, calculated at the LNO-CCSD level with "normal" convergence criteria. This is denoted as LNO-CCSD/def2-{T,Q}ZVPP for short.

(b) The (T) component was computed at the LNO-CCSD(T)/def2-{SVP,TZVPP} level within tight+ convergence criteria. (The reason for this choice will be explained below.)

(c) The difference between canonical CCSD(T) and LNO-LCCSD(T) was evaluated with the def2-SVP basis set, the largest for which we were able to carry out canonical

**Table 1. Forward and Reverse Barrier Heights (kcal/mol) for the Modified MOBH35 Dataset[a]**

| | $V_f^{\ddagger}, \Delta E^{\#}_{fwd}$ | | | | | $V_r^{\ddagger}, \Delta E^{\#}_{rev}$ | | | | |
|---|---|---|---|---|---|---|---|---|---|---|
| rxn | CCSD(T)/def2-SVP | $\Delta E_{(DLPNO-CCSD(T1) - CCSD(T)/def2-SV(P))}$ | $\Delta E_{(DLPNO-CCSD(T1) - CCSD(T)/def2-SVP)}$ | Iron and Janes | best est. in this work[c] | CCSD(T)/def2-SVP | $\Delta E_{(DLPNO-CCSD(T1) - CCSD(T)/def2-SV(P))}$ | $\Delta E_{(DLPNO-CCSD(T1) - CCSD(T)/def2-SVP)}$ | Iron and Janes | best est. in this work[c] |
| **1** | 27.06 | 0.09 | 0.12 | 26.03 | 26.20 | 14.02 | 0.11 | 0.24 | 15.40 | 14.02 |
| **2** | 5.63 | −0.03 | −0.05 | 5.58 | 5.71 | 25.10 | 0.02 | 0.04 | 22.11 | 22.25 |
| **3** | 0.95 | 0.05 | 0.01 | 0.91 | 0.92 | 27.07 | −0.47 | −0.55 | 27.21 | 26.92 |
| **4** | 2.36 | 0.12 | −0.07 | 1.49 | 1.36 | 8.60 | −0.11 | −0.24 | 8.86 | 8.25 |
| **5** | 4.68 | −0.65 | −0.67 | 4.47 | 4.63 | 22.02 | 0.03 | 0.12 | 22.76 | 22.60 |
| **6** | 13.44 | −0.07 | 0.10 | 15.77 | 15.76 | 13.56 | −0.78 | −0.62 | 14.25 | 14.61 |
| **7** | 26.66 | −0.12 | 0.09 | 27.94 | 27.59 | 18.29 | −1.00 | −0.86 | 18.47 | 18.58 |
| **8** | 36.92 | 3.68 | 3.93 | 37.28 | 34.57 | 32.30 | 2.70 | 2.97 | 35.82 | 31.82 |
| **9** | 28.59 | 3.42 | 3.59 | 33.00 | 27.71 | 15.20 | −7.33 | −7.49 | 4.93 | 11.97 |
| **10** | −3.48 | −0.84 | −0.83 | −5.28 | −4.29 | 9.58 | 0.17 | 0.18 | 7.67 | 8.22 |
| **11** | 29.81 | −0.63 | −0.38 | 29.90[b] | 29.49 | 84.09 | 0.63 | 0.60 | 84.70[b] | 82.34 |
| **12** | 5.67 | −0.15 | −0.18 | 5.04[b] | 5.50 | 36.83 | −0.27 | −0.18 | 36.69[b] | 37.18 |
| **13** | 18.37 | 1.81 | 1.80 | 22.41 | 20.65 | 48.18 | 1.27 | 1.29 | 49.69 | 47.99 |
| **14** | 10.17 | 0.20 | 0.15 | 10.35[b] | 10.10 | 13.38 | 0.00 | −0.16 | 13.67[b] | 14.37 |
| **15** | 23.90 | −0.11 | 0.11 | 20.27 | 20.66 | 74.84 | 0.28 | 0.45 | 77.23 | 74.98 |
| **16** | 37.45 | −0.82 | −0.76 | 34.22 | 35.45 | 55.56 | 0.92 | 1.03 | 55.40 | 53.77 |
| **21** | 11.11 | 0.26 | 0.48 | 9.18 | 8.41 | 11.11 | 0.20 | 0.44 | 9.20 | 8.41 |
| **22** | 14.86 | −0.08 | 0.04 | 14.30 | 13.84 | 30.88 | 0.56 | 0.69 | 29.05 | 27.01 |
| **23** | 29.48 | 0.85 | 0.70 | 30.71 | 29.45 | 20.80 | 0.54 | 0.43 | 21.19 | 20.35 |
| **26** | 21.92 | 0.27 | 0.28 | 25.39 | 25.83 | −0.07 | −0.01 | 0.00 | 0.19 | 0.11 |
| **27** | 16.09 | −0.03 | 0.06 | 13.76 | 14.05 | 1.29 | 0.04 | 0.05 | 2.39 | 2.29 |
| **28** | 31.96 | −0.52 | −0.32 | 29.06 | 30.18 | 16.85 | 0.54 | 0.51 | 16.63 | 15.52 |
| **29** | 15.74 | 0.20 | 0.21 | 14.95 | 14.72 | 33.86 | −0.26 | −0.13 | 30.88 | 31.19 |
| **30** | 10.87 | −0.04 | −0.01 | 9.88 | 9.79 | 19.88 | −0.04 | −0.13 | 17.22 | 16.60 |
| **31** | 2.14 | 0.18 | 0.32 | 3.25 | 2.91 | 12.37 | 0.00 | −0.09 | 13.33 | 12.90 |
| **32** | 23.66 | −0.59 | −0.56 | 19.16 | 20.18 | 58.44 | 0.38 | 0.35 | 64.56 | 62.62 |
| **33** | 2.77 | 0.04 | 0.03 | 1.26 | 1.05 | 9.96 | −0.08 | −0.07 | 7.83 | 7.86 |
| **34** | 28.85 | −0.15 | −0.21 | 29.15 | 29.16 | 4.32 | −0.06 | −0.05 | 2.91 | 3.04 |
| **35** | 14.97 | 0.69 | 0.63 | 18.31 | 17.28 | −3.85 | 0.19 | 0.24 | −1.41 | −2.44 |

[a]CCSD(T)/CBS$_{W1}$+Δ(T)/TZVP[a]. Calculated values obtained from ref 34. "CCSD(T)/CBS$_{W1}$" denotes an analogous extrapolation to that employed in W1[98] for the triples, CCSD, and SCF energies, abbreviated similarly as W[ST,TQ,TQ], where S, T, and Q denote the def2-SVP, def2-TZVPP (only a single set of polarization functions was used for the triples term), and def2-QZVPP basis sets. "Δ(T)/TZVP" is the [DLPNO-CCSD($T_1$) − DLPNO-CCSD($T_0$)] difference in a def2-TZVP basis set, but we will also make use of the less confusing $T_1$−$T_0$ symbolism for that term; $T_1$ stands for an iterative treatment of the triples terms in DLPNO-CCSD($T_1$), whereas $T_0$ refers to a semicanonical perturbative treatment of the same term in DLPNO-CCSD($T_0$). [b]Modified values at revised geometries from Dohm et al.[45] These energies come from the PWPB95-D4/def2-{T,Q}ZVPP level of theory from the modified MOBH35 article. [c]Best estimated energies calculated in this work (see the text for discussion).

calculations on all systems other than the bimolecular reactions **17−20**.

The final computed results are given in Table 1, compared with the previous computed reference values of Iron and Janes. The latter were obtained at what those authors term an approximation to W1 theory, namely, DLPNO-CCSD(T)/def2-{T,Q}VPP + [DLPNO-CCSD($T_1$) − DLPNO-CCSD($T_0$)]/def2-TZVP.

For the most part, differences between old and new reference values are fairly modest, with a 1.70 kcal/mol RMS (root mean square). However, there are two outliers: reactions **8** and, especially, **9**, which are really two consecutive steps of the same reaction (Scheme 1).

For reaction **9**, discrepancies between old and new values reach a startling 5.3 and 7.0 kcal/mol on the forward and reverse barriers, respectively; for reaction **8**, the 2.7 kcal/mol value for the forward reaction is paralleled by 1.8 kcal/mol for reaction **13**, and 1.7 kcal/mol for the reverse reaction comes quite close to the largest errors of reverse reactions **15**, **16**, and **32**.

If we delete these two reactions, then the RMSD (root-mean-square difference) between the two sets of best estimates (Iron−Janes and present work) drops from 2.41 to 1.68 kcal/mol; if we only delete reaction **9**, then the RMSD is 1.73 kcal/mol. For overall reaction energies $\Delta E_r = \Delta E^{\#}_{fwd} - \Delta E^{\#}_{rev}$, the errors in reaction **9** mutually amplify, leading to an alarming 12.3 kcal/mol discrepancy between the two sets of numbers. (In reaction **8**, this drops to just 1.3 kcal/mol owing to a felicitous mutual cancellation.)

**Scheme 1. Consecutive Reactions 8 and 9 in the MOBH35 Database**

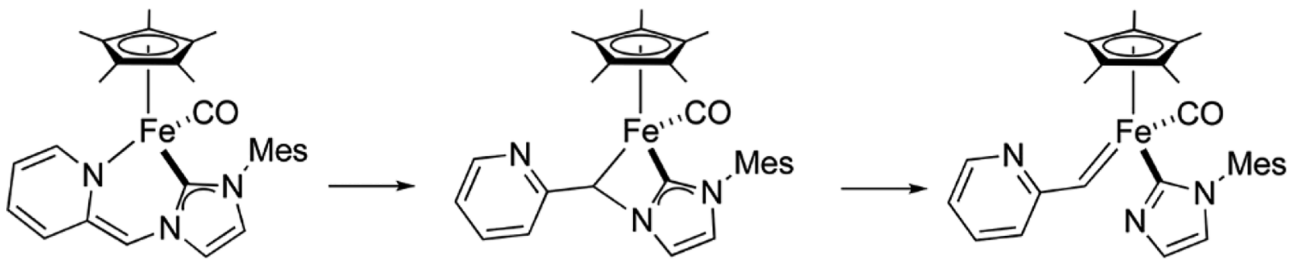

Why is such a thing happening? In a previous study by Sylvetsky et al.[38] on isomerism in polypyrroles, it was reported that the accuracy of the DLPNO-CCSD(T) and, to a lesser extent, DLPNO-CCSD($T_1$) methods suffers for systems that have significant static correlation—in that case, the Möbius rings and the transition states leading to them. (The Hückel and figure-eight isomers, in contrast, were much easier for all localized approaches.)

Table 2. Multireference Diagnostics and Energy Differences (kcal/mol) for Selected Reactions in MOBH35[a]

| | $T_1$ | $D_1$ | $D_2$ | FOD (TPSS) | FOD (TPSS0) | $\max\|t_i^A\|$ | $\max\|t_{ij}^{AB}\|$ | %$E_{corr}$[(T)] | $I_{ND}/I_{tot}$ | rev. $I_{ND}$ | $M$ | H-L gap (eV) | ΔE(DLPNO-CCSD($T_1$) – CCSD(T)) | ΔE(PNO-LCCSD(T) – CCSD(T)) | ΔE(LNO-CCSD(T) – CCSD(T)) |
|---|---|---|---|---|---|---|---|---|---|---|---|---|---|---|---|
| $P_6$ | 0.040 | 0.204 | 0.196 | 0.509 | 0.843 | 0.073 | 0.043 | 4.425 | 0.217 | 0.016 | 0.079 | 9.884 | -0.078 | 0.196 | -0.069 |
| $R_6$ | 0.039 | 0.201 | 0.190 | 0.419 | 0.747 | 0.104 | 0.025 | 4.371 | 0.217 | 0.016 | 0.076 | 10.605 | -0.790 | -0.005 | -0.265 |
| $TS_6$ | 0.039 | 0.195 | 0.190 | 0.413 | 0.759 | 0.100 | 0.037 | 4.441 | 0.218 | 0.016 | 0.076 | 10.194 | 0.712 | 0.201 | 0.196 |
| $P_7$ | 0.042 | 0.223 | 0.196 | 0.667 | 0.979 | 0.111 | 0.029 | 4.439 | 0.218 | 0.016 | 0.079 | 9.986 | -0.118 | 0.098 | -0.239 |
| $R_7$ | 0.040 | 0.204 | 0.196 | 0.509 | 0.843 | 0.073 | 0.043 | 4.425 | 0.217 | 0.016 | 0.079 | 9.884 | -0.989 | -0.041 | -0.245 |
| $TS_7$ | 0.041 | 0.206 | 0.193 | 0.504 | 0.888 | 0.115 | 0.068 | 4.584 | 0.221 | 0.016 | 0.081 | 9.780 | 0.870 | 0.139 | 0.006 |
| $P_8$ | 0.034 | 0.242 | 0.202 | 0.417 | 1.002 | 0.098 | 0.075 | 4.690 | 0.253 | 0.017 | 0.107 | 10.570 | 3.667 | 3.102 | 0.488 |
| $R_8$ | 0.033 | 0.232 | 0.198 | 0.694 | 1.306 | 0.082 | 0.049 | 4.719 | 0.252 | 0.017 | 0.101 | 8.631 | 2.698 | 3.247 | 0.186 |
| $TS_8$ | 0.037 | 0.321 | 0.216 | 1.176 | 1.551 | 0.094 | 0.047 | 4.719 | 0.250 | 0.016 | 0.098 | 8.769 | 0.969 | -0.145 | 0.302 |
| $P_9$ | 0.050 | 0.513 | 0.296 | 0.823 | 1.470 | 0.194 | 0.066 | 4.949 | 0.256 | 0.017 | 0.117 | 7.385 | 3.419 | 1.096 | 0.530 |
| $R_9$ | 0.034 | 0.242 | 0.202 | 0.419 | 1.002 | 0.098 | 0.075 | 4.690 | 0.253 | 0.017 | 0.107 | 10.570 | -7.327 | -1.939 | -0.951 |
| $TS_9$ | 0.039 | 0.331 | 0.219 | 0.578 | 1.218 | 0.137 | 0.065 | 4.782 | 0.255 | 0.017 | 0.107 | 9.281 | 10.746 | 3.036 | 1.481 |
| $P_{10}$ | 0.014 | 0.076 | 0.174 | 0.217 | 0.760 | 0.035 | 0.013 | 3.696 | 0.223 | 0.015 | 0.105 | 9.003 | -0.840 | -0.521 | -0.292 |
| $R_{10}$ | 0.018 | 0.108 | 0.228 | 0.899 | 1.522 | 0.061 | 0.114 | 3.871 | 0.227 | 0.015 | 0.138 | 6.301 | 0.172 | 0.041 | 0.075 |
| $TS_{10}$ | 0.015 | 0.080 | 0.180 | 0.316 | 0.892 | 0.043 | 0.023 | 3.819 | 0.227 | 0.015 | 0.101 | 8.814 | -1.012 | -0.562 | -0.367 |
| $P_{13}$ | 0.021 | 0.136 | 0.188 | 0.545 | 1.155 | 0.061 | 0.033 | 4.470 | 0.241 | 0.017 | 0.094 | 8.302 | 1.806 | 0.365 | -0.071 |
| $R_{13}$ | 0.020 | 0.108 | 0.211 | 0.646 | 1.320 | 0.054 | 0.063 | 4.497 | 0.242 | 0.017 | 0.110 | 8.345 | 1.273 | 0.433 | 0.113 |
| $TS_{13}$ | 0.023 | 0.139 | 0.202 | 0.871 | 1.521 | 0.067 | 0.043 | 4.682 | 0.246 | 0.018 | 0.107 | 7.673 | 0.533 | -0.068 | -0.184 |
| $P_{15}$ | 0.025 | 0.099 | 0.175 | 0.282 | 0.650 | 0.058 | 0.032 | 4.572 | 0.207 | 0.016 | 0.083 | 9.510 | -0.113 | 0.136 | -0.083 |
| $R_{15}$ | 0.035 | 0.195 | 0.219 | 0.566 | 1.044 | 0.148 | 0.060 | 5.015 | 0.217 | 0.018 | 0.124 | 8.069 | 0.279 | 0.322 | -0.088 |
| $TS_{15}$ | 0.030 | 0.152 | 0.210 | 0.663 | 1.124 | 0.100 | 0.061 | 5.143 | 0.218 | 0.018 | 0.119 | 8.636 | -0.392 | -0.186 | 0.005 |
| $P_{16}$ | 0.018 | 0.082 | 0.208 | 0.898 | 1.125 | 0.056 | 0.087 | 4.842 | 0.211 | 0.017 | 0.106 | 8.885 | -0.824 | -0.327 | -0.099 |
| $R_{16}$ | 0.040 | 0.279 | 0.218 | 0.475 | 0.934 | 0.219 | 0.077 | 5.226 | 0.213 | 0.017 | 0.097 | 7.767 | 0.922 | -0.153 | 0.137 |
| $TS_{16}$ | 0.029 | 0.193 | 0.214 | 0.976 | 1.287 | 0.126 | 0.045 | 4.993 | 0.212 | 0.017 | 0.097 | 8.267 | -1.746 | -0.174 | -0.236 |
| $P_{23}$ | 0.016 | 0.101 | 0.206 | 0.474 | 1.062 | 0.065 | 0.060 | 4.527 | 0.249 | 0.018 | 0.160 | 8.947 | 0.854 | 0.344 | 0.089 |
| $R_{23}$ | 0.016 | 0.096 | 0.207 | 0.215 | 0.851 | 0.062 | 0.072 | 4.567 | 0.250 | 0.018 | 0.159 | 10.436 | 0.535 | -0.102 | 0.044 |
| $TS_{23}$ | 0.018 | 0.118 | 0.210 | 0.590 | 1.245 | 0.054 | 0.043 | 5.044 | 0.252 | 0.018 | 0.165 | 7.935 | 0.319 | 0.446 | 0.045 |
| $P_{28}$ | 0.017 | 0.092 | 0.186 | 0.428 | 1.052 | 0.044 | 0.066 | 3.964 | 0.229 | 0.016 | 0.086 | 8.632 | -0.519 | 0.206 | 0.128 |
| $R_{28}$ | 0.020 | 0.149 | 0.187 | 0.460 | 1.066 | 0.112 | 0.072 | 3.993 | 0.230 | 0.016 | 0.091 | 8.704 | 0.538 | 0.271 | -0.022 |
| $TS_{28}$ | 0.018 | 0.104 | 0.186 | 0.343 | 0.970 | 0.063 | 0.059 | 4.040 | 0.231 | 0.016 | 0.087 | 9.258 | -1.057 | -0.066 | 0.150 |
| $P_{32}$ | 0.017 | 0.066 | 0.181 | 0.223 | 0.482 | 0.034 | 0.051 | 4.337 | 0.196 | 0.015 | 0.081 | 9.855 | -0.592 | 0.055 | -0.032 |
| $R_{32}$ | 0.027 | 0.140 | 0.194 | 0.580 | 0.889 | 0.103 | 0.048 | 4.476 | 0.193 | 0.014 | 0.080 | 8.679 | 0.380 | 0.024 | 0.170 |
| $TS_{32}$ | 0.023 | 0.126 | 0.208 | 0.548 | 0.815 | 0.060 | 0.042 | 4.346 | 0.195 | 0.015 | 0.077 | 9.150 | -0.972 | 0.031 | -0.202 |

[a]Reactant species (R), products (P), and transition states (TS) in the MOBH35 database. Energy differences are all in a def2-SV(P) basis set. Heat mapping for diagnostics, within each column, is from green for the lowest to red for the highest, while for energy differences, it is from blue for the most negative via white for zero to red for the most positive. DLPNO-CCSD($T_1$) is within TightPNO settings, LNO-CCSD(T) is based on standard tight thresholds, and PNO-LCCSD(T) is also on tight settings. The very large errors for reaction **9** in DLPNO-CCSD($T_1$) are obvious and become even larger with the common DLPNO-CCSD($T_0$) approximation. Yet, the boxes for both approaches are comparatively narrow: the distribution is strongly leptokurtic (long-tailed), and reactions **9**, **8**, and to a lesser extent **16** are "extreme outliers", outliers beyond 3 IQR.

Table 2 presents a large number of diagnostics for static correlation for selected structures presently under study: the full set can be found in the SI.

For product **9**, $T_1$ = 0.05, $D_1$ = 0.52, $D_2$ = 0.30, and $\max|t_i^A|$ = 0.19 all indicate severe static correlation. Moreover, these indicators monotonically increase from reactant **9** via $TS_9$ to product **9**; hence, no error compensation can reasonably be hoped for. (We also note that the HOMO−LUMO gap steadily narrows from 10.6 via 9.3 to 7.4 eV.) In fact, in the earliest communication on DLPNO-CCSD($T_1$) states, it was expected then that the $T_0$ approximation would fail in rare cases of small $\Delta E_{H\text{-}L}$ gaps.[36] In contrast, these indicators, while elevated, are similar for reactant **8** and product **8** (≡reactant **9**), which may explain the observed error compensation in the reaction energy. (However, $\Delta E_{H\text{-}L}$ alone clearly does not tell the whole story, as evidenced by the even smaller $\Delta E_{H\text{-}L}$ in reactant **10**, where, however, $T_1$, $D_1$, and $\max|t_i^A|$ are much smaller than those for product **9**. We note that in other situations where both the singles- and doubles-based diagnostics are elevated, like in reaction **16**, the same problem is experienced in a milder form.) We also ought not to lose sight of the fact that the CCSD(T) approximation itself will become problematic for sufficiently strong static correlation, and hence, its accurate reproduction under such circumstances may be a somewhat misguided target.

A standard Tukey box plot[99] of errors in all MOBH35 forward and reverse barriers, as well as reaction energies, is given in Figure 1; see also Table S2 in the SI for the percentile summaries in a numerical form. As always, the box indicates the IQR or interquartile range, the distance between the 25th and 75th percentile, and the dividing line in the box the median of the distribution. The whiskers were defined in the most commonly used standard manner: from the box to the last data point that is still within 1.5 IQR from it. Points outside are indicated as outliers (if between 1.5 and 3 IQR from the box) and extreme outliers (if further than 3 IQR away). For a normal distribution, the whiskers would be symmetric and encompass ±2.698$\sigma$ or 99.3% of all data points.

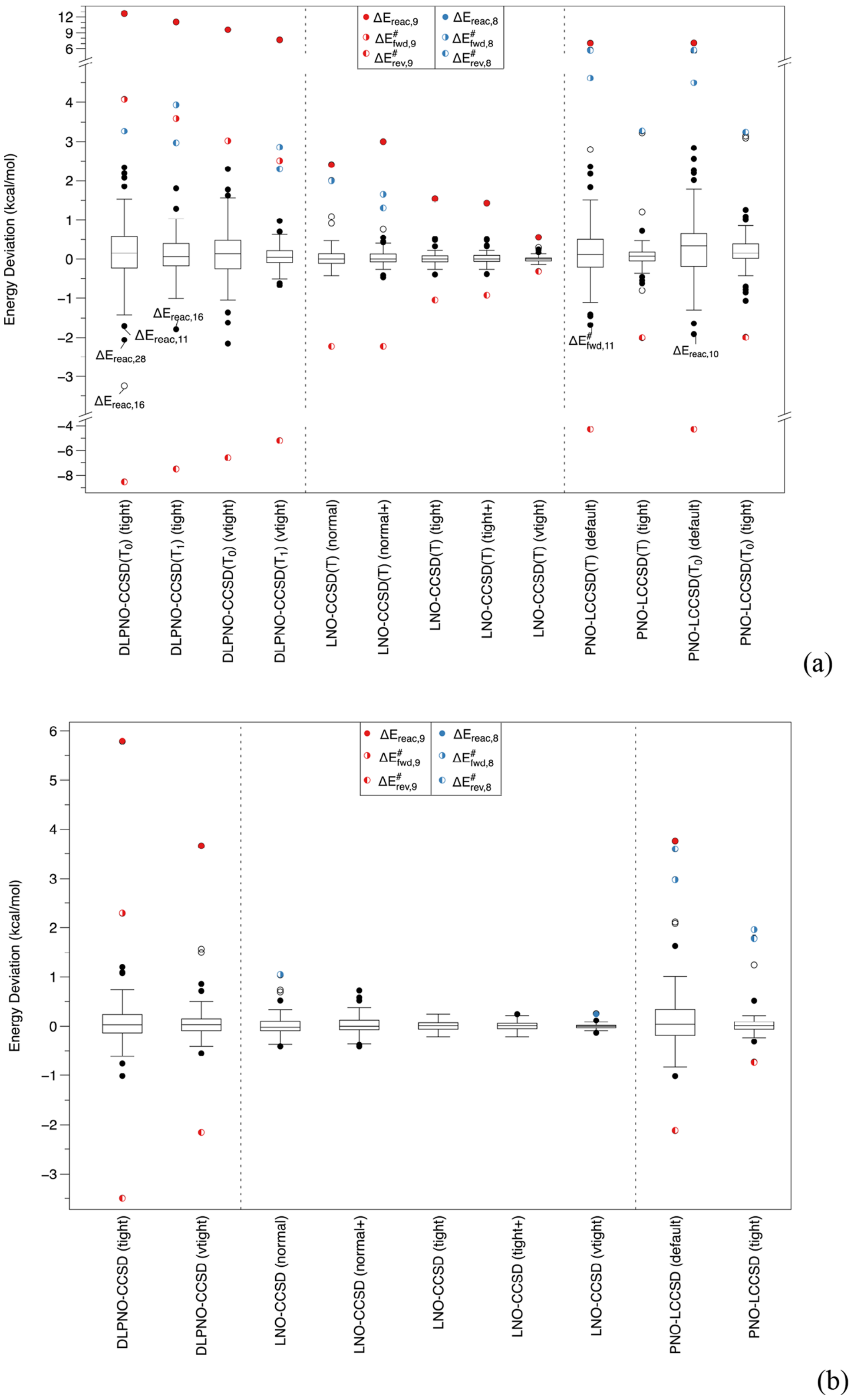

**Figure 1.** Box-and-whisker plot for the energy deviations of (a) local CCSD(T) approximations from canonical CCSD(T); (b) local CCSD approximations from canonical CCSD. The def2-SVP basis set was used throughout. Pair tolerances for LNO-CCSD(T) are as follows: normal (wpairtol = $1 \times 10^{-5}$), normal+ (wpairtol = $1 \times 10^{-6}$), tight (wpairtol = $3 \times 10^{-6}$), and tight+ and vTight (both with wpairtol = $1 \times 10^{-6}$).

Our largest canonical calculations (reactions **8** and **9**) took several weeks each on Intel Skylake and Cascade Lake machines with 36 and 40 cores, respectively, and 384 GB of RAM. Clearly, repeating this for def2-TZVPP is not going to be a realistic option with the available hardware. However, would canonical calculations with the even smaller def2-SV(P) basis have been equally informative? As can be seen in Figure S1 in the Supporting Information (SI), the box plot looks extremely

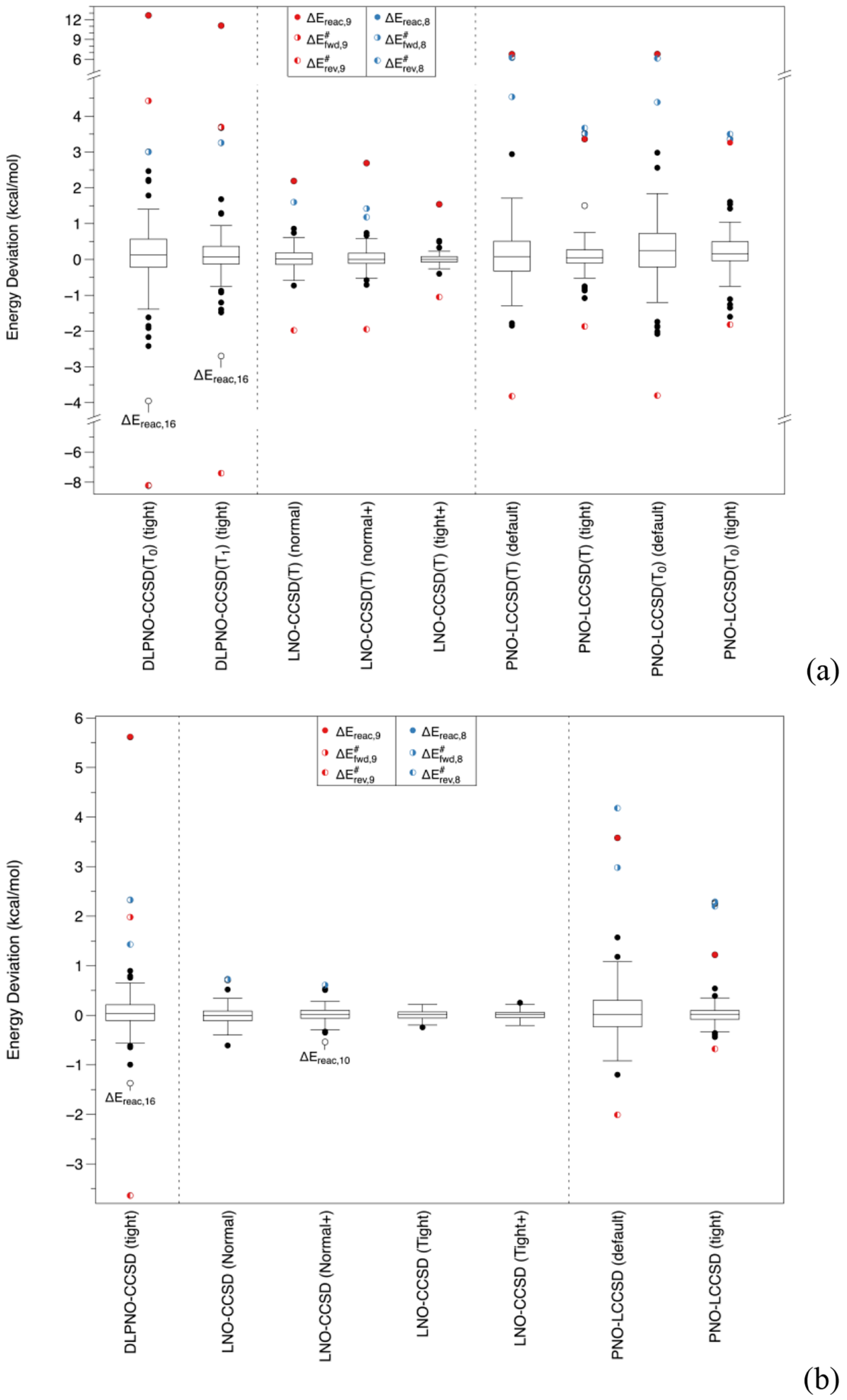

**Figure 2.** (a,b) Same comparison as in Figure 1 but now for a def2-TZVPP basis set. The canonical reference numbers were obtained by a composite model approximation (see eqs 1 and 2).

similar. This may be useful for future calibration work if a canonical reference is required.

As we can see in Figure 1, this is not purely a problem with the connected triples: reaction **9** is an outlier for the DLPNO-CCSD vs CCSD difference as well. The boxes have similar widths between DLPNO-CCSD and DLPNO-CCSD($T_1$), while that of DLPNO-CCSD($T_0$) is about double the width. This implies that the more economical $T_0$ approximation comes at a significant price in accuracy even for "nonoutlier" systems.

Ma and Werner's PNO-LCCSD with default cutoffs has a much wider box than DLPNO-CCSD with tight cutoffs, but interestingly, the outliers (reaction **9** and the reverse barrier of **8**) are much less far off. The boxes of PNO-LCCSD and PNO-LCCSD(T) (the latter both with and without the $T_0$ approximation) have similar widths—just the outliers more spread out with than without the triples. Still, PNO-LCCSD(T) with tight criteria does very well for systems other than the outliers.

LNO-CCSD even with "normal" cutoffs appears to be still more resilient to outliers than PNO-LCCSD "tight"; for LNO-CCSD(T), performance with normal criteria is about the same quality as PNO-LCCSD(T) "tight".

In ref 38, the best results were obtained for LNO-CCSD(T) with "tight" criteria, and in addition, wpairtol tightened further to $10^{-6}$ (the value from "vTight")—we denote this combination "tight+", which yielded almost the same accuracy as "vTight" but at considerably reduced cost. However, in that regard, the very extended $\pi$ systems (polypyrroles) of ref 38 are very different from the systems at hand, where "tight+" and standard "tight" yield fundamentally the same error distribution for CCSD(T). (For CCSD, tight+ has a narrower IQR, 0.10 kcal/mol, than tight, 0.13 kcal/mol.) At the expense of increasing CPU time by a factor of about 3–4, "tight" yields a quite narrow error distribution (IQR = 0.15 for LNO-CCSD(T)); we still have two outliers for reaction **9**, but they are now in the 1 kcal/mol range, while the whisker span is comparable to PNO-LCCSD(T) tight.

Would tightening LNO criteria further eventually lead to convergence to the canonical results? One could set all cutoffs to zero, but then, one would essentially have a clumsy way to do canonical CCSD(T). Instead, we used the "vTight" or "veryTight" cutoffs throughout, which increase our computational cost by about a further factor of three for the largest systems. However, the payoff was an IQR for LNO-CCSD$_{\text{vTight}}$ of just 0.05 kcal/mol (!) and for LNO-CCSD$(T)_{\text{vTight}}$ of just 0.07 kcal/mol. True, LNO-CCSD$(T)_{\text{vTight}}$ has an "extreme" outlier of 0.56 kcal/mol, but that is more of a testimony to the very small IQR than to any sort of breakdown of the approximation.

We conclude that of all the localized orbital-based approaches, LNO-CCSD$(T)_{\text{vTight}}$ is the most resilient to systems with severe static correlation and can be used as an alternative to canonical CCSD(T) if the latter is simply beyond available computational resources. This is especially true for larger basis sets where the canonical calculation would be completely intractable.

Would tightening wpairtol by itself to the "Tight" value, and otherwise using "Normal" cutoffs, be adequate? From Figure 1a, this seems to rein in some outliers in LNO-CCSD, but from Figure 1b, it is clear that this does not help for LNO-CCSD(T).

One of the original purposes of MOBH35 was to use the data to evaluate DFT functionals for organometallic and catalysis applications. Given the way that localized coupled cluster methods struggle to cope with the static correlation in reaction **9**, one might legitimately wonder if canonical CCSD(T) is itself adequate for this reaction and whether it can still be considered a reasonable test for any DFT method.

We checked the effects of deleting reaction **9** on the performance statistics of various functionals in Iron & Janes. The comparison is given in the SI for a number of DSD3 and other common functionals and is based on the revised reference energies reported in this work; for some newer functionals not covered in Iron & Janes, we would like to refer the reader to Figure 3 in both refs 100 and 101. The bottom line is that while MAD drops slightly for many functionals, it does not affect any conclusions about their relative accuracy; for the best double hybrids, however, one may wish to exercise caution. (For reasons related to basis set superposition error rather than static correlation, one may wish to eliminate the bimolecular reactions **17–20**, unless one is prepared to carry out at least a {T,Q} basis set extrapolation or apply the counterpoise method.)

It may be argued that def2-SVP is too small as a basis set for the various methods to be at their best. Unfortunately, a comparison against canonical CCSD(T)/def2-TZVPP is simply not feasible for the systems of greatest interest, but an approximate set of CCSD(T)/def2-TZVPP reference values can be obtained by applying the additivity approximation

$$\begin{aligned} &E[\text{CCSD(T)/def2} - \text{TZVPP}] \\ &\quad\approx E[\text{LNO} - \text{CCSD(T)/def2} - \text{TZVPP}_{\text{Tight+}}] \\ &\quad\quad - E[\text{LNO} - \text{CCSD(T)/def2} - \text{SVP}_{\text{Tight+}}] \\ &\quad\quad + E[\text{CCSD(T)/def2} - \text{SVP}] \end{aligned} \tag{1}$$

and, similarly,

$$\begin{aligned} &E[\text{CCSD/def2} - \text{TZVPP}] \\ &\quad\approx E[\text{LNO} - \text{CCSD/def2} - \text{TZVPP}_{\text{Tight+}}] \\ &\quad\quad - E[\text{LNO} - \text{CCSD/def2} - \text{SVP}_{\text{Tight+}}] \\ &\quad\quad + E[\text{CCSD/def2} - \text{SVP}] \end{aligned} \tag{2}$$

This graph is presented in Figure 2a,b for CCSD(T) and CCSD, respectively, and the corresponding numerical values are shown in Table S1 in the SI. With some quantitative changes, the same basic conclusions apply, and the box plot widths and outliers are similar, though errors of PNO and DLPNO approaches appear to be narrowed.

This implies that our observations about the performance of the various localized CCSD and CCSD(T) approaches are fairly basis set-independent. This would actually make sense if the differences are largely driven by type A static correlation (as Hollett and Gill[83] denote absolute near-degeneracy correlation). However, as shown by Karton et al.,[88] the higher one goes in the coupled cluster hierarchy beyond CCSD(T), the more rapidly their contributions converge with the basis set to the point that, for example, quintuple excitations are fully converged even with an unpolarized (!) double-zeta basis set (as we showed there, e.g., for ozone and $C_2$).

As even with normal cutoffs, LNO-CCSD(T) performs surprisingly well for the TZVPP basis set; we have elected to use LNO-CCSD$(T)_{\text{Normal}}$/def2-{T,Q}ZVPP for the CBS extrapolation of the correlation energy. The HF component was extrapolated separately. (See the Computational Methods section for extrapolation details.)

For the (T) component, on the other hand, we carried out LNO-CCSD$(T)_{\text{Tight+}}$/def2-{SVP,TZVPP} extrapolation.

As the third tier of our composite "best estimate", we use the difference between canonical CCSD(T)/def2-SVP and LNO-CCSD$(T)_{\text{Tight+}}$/def2-SVP.

A comparison between our best estimates thus obtained, and the older values of Iron and Janes, can be found in Table 1. Only for two reactions, **8** and **9**, are there discrepancies significantly in excess of 2 kcal/mol; additional discrepancies at or near 2 kcal/mol are found for reactions **11**, **13**, **16**, **22**, and **32**. Let us now consider the diagnostics for these reactions in more detail.

For reaction **8**, we see $D_1$ to be similar between the reactant and the product but elevated for the transition state, and likewise for $T_1$. The FOD(TPSS) is likewise elevated for the transition state. This is consistent with the observation that the forward and reverse barriers differ quite significantly both between the two sets of best estimates, and between DLPNO-CCSD($T_1$) and canonical CCSD(T), but that for the reaction energy, the errors largely cancel.

Conversely, for reaction **9** (the reaction of which is the product of **8**), we see a steady increase in $T_1$, $D_1$, and indeed in max$|t_i^A|$ from the reactant via the transition state to the product. Moreover, the reciprocal HOMO−LUMO gap increases in tandem; all of this reflects steadily increasing type A static correlation. Not only are there large discrepancies between the two sets of best estimates for both barrier heights, but the errors actually amplify each other in the reaction energy.

Reactions **16**, and to a lesser extent **15**, have milder versions of the same scenario but in reverse: the $T_1$, $D_1$, and max$|t_i^A|$ diagnostics all decrease from the reactant to the transition state to the product, in tandem with the reciprocal HOMO−LUMO gap. Here too, discrepancies for forward and reverse barriers amplify each other in the reaction energy.

Furthermore, while reaction **32** has less multireference character, we see the same progression as in reactions **15** and **16** and hence similar observations about discrepancies.

At the request of a reviewer, an examination of the percentage of correlation energy recovered was carried out, and for DLPNO-CCSD($T_1$), the following additional accuracy settings aside from NormalPNO and TightPNO were considered: VeryTightPNO as detailed in the Computational Methods section, TightPNO with TCutPNO = $10^{-6}$ $E_h$, and TightPNO with $10^{-8}$ $E_h$ (10× looser and tighter, respectively, than the default). The latter was also employed in the 2-point {6,7} and {7,8} TCutPNO extrapolation formulas for the DLPNO-CCSD($T_1$) correlation energy proposed by Altun et al.[102] The percentages of correlation energy recovered ($E_{corr}$) are depicted in Figure 3, and the corresponding RMSD values are listed in Table S6.

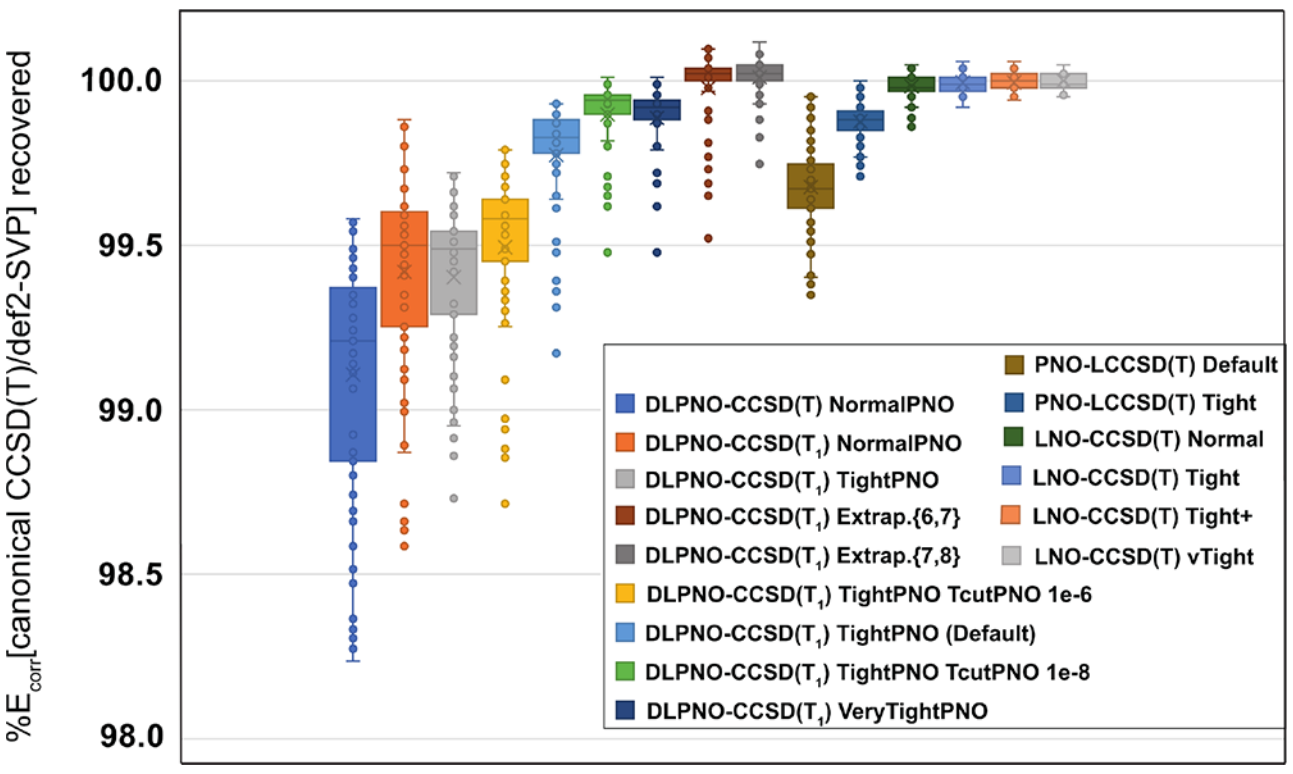

**Figure 3.** Percentages of correlation energy recovered from canonical CCSD(T)/def2-SVP with different accuracy settings in local coupled cluster approaches.

First, the percentage of $E_{corr}$ recovered is clearly sensitive to the degree of static correlation in all cases, especially for DLPNO-CCSD($T_1$), where even veryTightPNO is not immune to significant errors for reactions **8** and **9**.

Second, TightPNO DLPNO-CCSD(T1) and Tight PNO-LCCSD(T) both have a median %$E_{corr}$ recovery of around 99.8%, although the former has both a wider "box" (larger interquartile range, IQR) and more outliers. With {6,7} extrapolation, which aside from TightPNO DLPNO-CCSD($T_1$) requires an additional calculation at roughly NormalPNO cost, the IQR is much reduced and the median lies near 100%, similar to the LNO-CCSD(T) calculations in both aspects. Significant "under-recovered" outliers remain, however, particularly for reactions **8** and **9**. {7,8} Extrapolation, which requires calculations at TightPNO and VeryTightPNO cost, mitigates the outliers but does not improve the IQR. Even {6,7} extrapolation appears to be superior to VeryTightPNO on its own for this application.

For the LNO-CCSD(T) series, even Normal has a remarkably narrow spread, aside from the outliers for reactions **8** and **9**; the 1.5 IQR "whiskers" span a range comparable to {6,7} DLPNO extrapolation. LNO-CCSD(T) Tight has whiskers of similar width to Normal but no more outliers outside them. It is not clear that further gains from vTight are commensurate with the great additional computational cost for the present application. (This latter situation is apparently quite different for accurate benchmarks of noncovalent interactions.[103])

**A Remark on Static Correlation Diagnostics.** In order to get our bearings concerning static correlation diagnostics, we obtained them for a set of small closed-shell molecules, with all systems under investigation here being closed-shell as well and open-shell versions of some diagnostics not being uniquely defined. Lee[81] discussed this issue for the $T_1$ and $D_1$ diagnostics and proposed an open-shell generalization of the latter, while no published extension of $D_2$ to open-shell systems exists anywhere (MOLPRO, PSI4, and TURBOMOLE use different ad hoc versions).

We took the subset of 160 closed-shell species in the W4−17 dataset as our reference sample. Many of the diagnostics were extracted from the SI of ref 104; others were newly obtained for this work.

Subsequently, principal component analysis (PCA) on the correlation matrix and variable clustering analysis were carried out in JMP Pro 16. The first four principal components are given in Table 3; variables are grouped by the four clusters from the cluster analysis.

The separation between the two clusters designated as 2a and 2b is somewhat weak, but clusters 1 and 3 behave quite distinctly from each other and from 2a/2b. In a PCA on the correlation matrix, the eigenvalues should add up to the total number of variables; we can thus see that just the first two PCs contain the same information as 14.45 variables out of 18, and the first four the same as 16.61 variables. In PC1 all four clusters move in parallel, while in PC2 clusters 1 and 3, they move opposite to clusters 2a and 2b.

Diagnostics based on fractions of the total atomization energy are really intended for small molecules and become a bit unwieldy for, say, a molecule with 67 atoms (like in reactions **8** and **9**). Moreover, post-CCSD(T) corrections are quite simply unrealistic here. The results of a PCA (and cluster analysis) on such diagnostics as we were able to evaluate for MOBH35 are given in Table 4. (The correlation matrix can be found in Table S7 in the SI.)

There are some differences with the behavior of the diagnostics for W4−17, but a few things stand out: notably, both the W4−17 and MOBH35 diagnostic collections have a clear "principal component of strong correlation", and the difference between DLPNO-CCSD($T_1$) and DLPNO-CCSD($T_0$) is closely tied to the degree of static correlation.

**Table 3. Principal Component Analysis (Correlation Matrix) and Variable Clustering Analysis on the Subset of 160 Closed-Shell Molecules in the W4−17 Small-Molecule Thermochemistry Benchmark**[a]

| Cluster | Diagnostic | PC1 | PC2 | PC3 | PC4 | PC5 |
|---|---|---|---|---|---|---|
| 1 | $T_1$ | 0.25 | 0.05 | -0.45 | 0.10 | -0.19 |
| | $D_1$ | 0.23 | 0.12 | -0.50 | -0.07 | -0.08 |
| | $\max\|t_i^A\|$ | 0.23 | 0.09 | -0.52 | -0.06 | -0.02 |
| 2a | $2\ln(A)/N_{val}$ | 0.24 | -0.27 | 0.00 | 0.29 | -0.14 |
| | $I_{ND}/(I_{ND}+I_D)$ | 0.27 | -0.18 | 0.09 | 0.18 | -0.23 |
| | $revI_{ND,2021}$ | 0.21 | -0.31 | 0.07 | 0.33 | -0.15 |
| | $S_{norm}$ | 0.19 | -0.35 | 0.17 | 0.27 | -0.15 |
| | $1/(\varepsilon_{LUMO}-\varepsilon_{HOMO})$ | 0.17 | -0.30 | 0.07 | -0.23 | 0.47 |
| 2b | FOD(TPSS) | 0.25 | -0.11 | -0.13 | 0.14 | 0.66 |
| | Truhlar M | 0.29 | -0.07 | 0.08 | -0.13 | 0.08 |
| | $D_2$ | 0.28 | -0.13 | 0.13 | -0.26 | -0.03 |
| | $\max\|t_{ij}^{AB}\|$ | 0.25 | -0.06 | 0.08 | -0.41 | 0.04 |
| | $-T_e$(SCF) | 0.24 | -0.07 | 0.15 | -0.55 | -0.36 |
| 3 | %TAE[(T)]AVTZ | 0.24 | 0.29 | 0.10 | 0.13 | 0.07 |
| | %TAE[T4] | 0.25 | 0.24 | 0.16 | 0.12 | 0.18 |
| | %TAE[corr]CCSD(T) | 0.21 | 0.35 | 0.21 | 0.10 | -0.02 |
| | %TAE[corr]TPSS | 0.21 | 0.34 | 0.21 | 0.10 | -0.01 |
| | A100[TPSS] | 0.20 | 0.36 | 0.17 | 0.06 | -0.05 |
| | Eigenvalues | 10.52 | 3.93 | 1.29 | 0.87 | 0.44 |
| | Variance (%) | 58.46 | 80.28 | 87.43 | 92.26 | 94.69 |

[a]Positive values appearing in progressively darker shades of blue, negative ones of red, and white for zero.

**Table 4. Eigenvectors of the First Five Principal Components of Static Correlation Diagnostics for the MOBH35 Set**[a]

| Cluster | Diagnostic | PC1 | PC2 | PC3 | PC4 | PC5 |
|---|---|---|---|---|---|---|
| *A* | $T_1$ | 0.26 | 0.31 | -0.28 | -0.14 | -0.01 |
| | $D_1$ | 0.30 | 0.17 | -0.31 | -0.17 | 0.02 |
| | $\max\|t_i^A\|$ | 0.28 | 0.21 | -0.27 | 0.00 | 0.09 |
| | $(T_1)/(T_0)$ | 0.07 | 0.45 | -0.19 | 0.13 | 0.25 |
| | $A_{100}$[TPSS] | 0.12 | 0.47 | 0.31 | 0.11 | -0.14 |
| | %TAE[ΔX] | 0.12 | 0.34 | 0.37 | 0.18 | 0.00 |
| *B* | FOD(PBE0) | 0.30 | -0.27 | -0.13 | 0.16 | -0.27 |
| | FOD(TPSS) | 0.29 | -0.12 | -0.16 | 0.35 | -0.30 |
| | 1/H-L gap | 0.26 | -0.22 | -0.11 | 0.42 | -0.14 |
| | $D_2$ | 0.32 | -0.05 | -0.02 | 0.14 | 0.15 |
| | $\max\|t_{ij}^{AB}\|$ | 0.16 | -0.21 | 0.13 | 0.35 | 0.68 |
| *C* | $I_{ND}/I_{tot}$ | 0.27 | -0.27 | 0.14 | -0.41 | -0.04 |
| | rev. $I_{ND}$ | 0.23 | -0.03 | 0.47 | -0.08 | -0.22 |
| | $\%E_{corr}$[(T)] | 0.30 | 0.11 | 0.34 | 0.02 | -0.12 |
| | *M* | 0.24 | -0.14 | 0.22 | -0.18 | 0.41 |
| | $\|(T_1)-(T_0)\|$ | 0.30 | -0.04 | -0.09 | -0.47 | 0.05 |
| | Eigenvalues | 7.10 | 2.85 | 1.87 | 1.15 | 0.77 |
| | Variance (%) | 44.40 | 62.22 | 73.93 | 81.13 | 85.96 |

[a]Positive values appearing in progressively darker shades of blue, negative ones of red, and white for zero.

Furthermore, as shown in Figure 4, that same difference appears to be correlated with the error relative to canonical CCSD(T).

In the above PCA, we identified the intercorrelations among diagnostics and grouped them into clusters of variables, but could we select a subset of *k* "principal variables" that would span a space similar to the first *k* principal components? This statistical problem is addressed by the "subselect" module developed by Cadima et al.[105] for the R statistical environment[106] (version 4.1.2). Using the "eleaps" algorithm within that

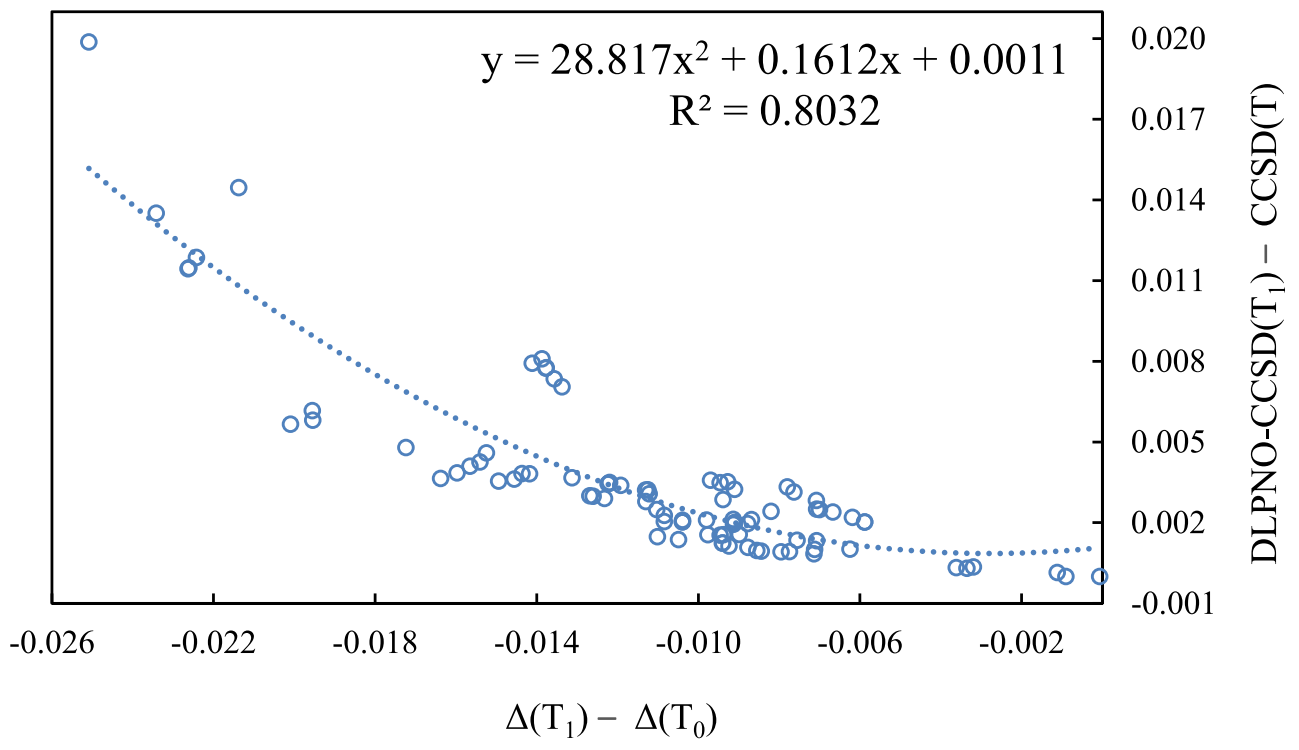

**Figure 4.** Correlation between the magnitude of $T_1$ correction to DLPNO-CCSD(T) and the residual discrepancy from canonical CCSD(T). Units are in kcal/mol, and the def2-SVP basis set was used throughout.

module with the GCD (generalized coefficient of determination) as the objective function, we obtained the following subsets of increasing size:

*k*=1: $D_2$ diagnostic, GCD=0.7128

*k*=2: $T_1$ + FOD(PBE0), GCD=0.7851

*k*=3: $T_1$ + $I_{ND}/I_{tot}$ + $A_{100}$[TPSS], GCD=0.8358 essentially tied with

$D_1$ + $I_{ND}/I_{tot}$ + $A_{100}$[TPSS], GCD=0.8351

*k*=4: $T_1$ + FOD(PBE0) + $I_{ND}/I_{tot}$ + $A_{100}$[TPSS], GCD=0.8848 just barely ahead of

$T_1$ + 1/gap + $I_{ND}/I_{tot}$ + $A_{100}$[TPSS], GCD=0.8805, slightly ahead of

$D_1$ + $(\varepsilon_{LUMO}-\varepsilon_{HOMO})^{-1}$ + $I_{ND}/I_{tot}$ + $A_{100}$[TPSS], GCD=0.8696

While we truncated the PCA after 4 PCs, we will list $k = 5$ for the sake of completeness:

*k*=5: $T_1$ + FOD(PBE0) + $I_{ND}/I_{tot}$ + $A_{100}$[TPSS] + $\max|t_{ij}^{AB}|$, GCD=0.8968, barely ahead of

$D_1$ + FOD(PBE0) + $I_{ND}/I_{tot}$ + $A_{100}$[TPSS] + $\max|t_{ij}^{AB}|$, GCD=0.8883

Let us now inspect the variables of the generated subsets and consider the clusters from the variable clustering analysis of the W4−17 diagnostics (Table 3). $D_2$ belongs to the cluster 2b subblock for a single variable, while for two variables, $T_1$ and FOD(PBE0) belong to clusters 1 and 2b, respectively. For $k = 3$, $I_{ND}/I_{tot}$ and $A_{100}$[TPSS] are representatives of clusters 2a and 3, respectively, while $T_1$ and $D_1$ both belong to cluster 1. For $k = 4$, we have $T_1$, $I_{ND}/I_{tot}$, FOD(PBE0), and $A_{100}$[TPSS] as one representative each of all four clusters: in that order, 1, 2a, 2b, and 3, respectively. Intriguingly, in the second solution, the reciprocal HOMO−LUMO gap $(\varepsilon_{LUMO} - \varepsilon_{HOMO})^{-1}$ belongs to cluster 2a.

**Timing Comparisons.** The calculations being reported above are run on a highly heterogeneous cluster, and de facto clock speeds vary because of differing workloads on the nodes and server cooling fluctuations. Therefore, we carried out separate timing calculations under controlled conditions.

Table 5 presents timings for a single system (the product of reaction **16**), all run on the same type of 16-core Intel Haswell servers with 256 GB RAM and 3.5 TB SSD scratch disk arrays. These servers were otherwise idle during the timing measurements. (The total memory and hard disk requirements (both in GB) can be found in Tables S4 and S5 in the SI, respectively.)

For such a small system, of course, canonical calculations are clearly faster except for the largest basis sets; needless to say, of course, for larger systems, canonical CCSD(T) CPU times will scale as $O(N^7)$, while all discussed localized orbital methods would scale asymptotically linearly.

Table 5. Wall-Clock Execution Times (h) for Local CC Single-Point Calculations for the Product of Reaction 16 of the MOBH35 Dataset on Two 8-Core Intel Xeon E5-2630 v3 CPUs (2.40 GHz)

| methods | threshold | def2-SV(P) | def2-SVP | def2-TZVP | def2-TZVPP | def2-QZVPP |
|---|---|---|---|---|---|---|
| DLPNO-CCSD(T) | NormalPNO | 0.05 | 0.06 | 0.27 | 0.41 | 2.23 |
| DLPNO-CCSD(T) | TightPNO[a] | 0.06 | 0.08 | 0.32 | 0.54 | 3.56 |
| DLPNO-CCSD(T) | TightPNO[b] | 0.15 | 0.20 | 0.94 | 1.40 | 6.24 |
| DLPNO-CCSD(T) | TightPNO[c] | 0.53 | 0.82 | 5.37 | 8.70 | 28.09 |
| DLPNO-CCSD(T) | veryTightPNO | 0.59 | 0.90 | 5.79 | 9.15 | 35.20 |
| $DLPNO-CCSD(T_1)$ | NormalPNO | 0.13 | 0.19 | 0.77 | 1.13 | 3.81 |
| $DLPNO-CCSD(T_1)$ | TightPNO[a] | 0.17 | 0.22 | 0.78 | 1.14 | 4.78 |
| $DLPNO-CCSD(T_1)$ | TightPNO[b] | 0.38 | 0.54 | 2.30 | 3.22 | 10.89 |
| $DLPNO-CCSD(T_1)$ | TightPNO[c] | 0.86 | 1.32 | 7.56 | 11.64 | 34.83 |
| $DLPNO-CCSD(T_1)$ | veryTightPNO | 0.92 | 1.43 | 7.93 | 12.36 | 43.38 |
| PNO-LCCSD(T) | Default | 0.13 | 0.15 | 0.81 | 1.14 | 3.75 |
| PNO-LCCSD(T) | Tight | 0.22 | 0.26 | 0.98 | 1.47 | 5.5 |
| LNO-CCSD(T) | Normal | 0.34 | 0.51 | 1.46 | 2.01 | 4.25 |
| LNO-CCSD(T) | Tight | 0.88 | 1.41 | 4.58 | 6.42 | 14.12 |
| LNO-CCSD(T) | vTight | 1.72 | 2.91 | 11.82 | 16.23 | 40.92 |
| canonical CCSD(T) | | 0.16 | 0.33 | 3.78 | 8.08 | 83.54 |
| Nbasis | | 184 | 221 | 410 | 506 | 958 |

[a]TightPNO with TcutPNO = $10^{-6}$ $E_h$. [b]TightPNO with TcutPNO = $10^{-7}$ $E_h$, i.e., the default TightPNO settings. [c]TightPNO with TcutPNO = $10^{-8}$ $E_h$.

The CPU time dependence on the size of the basis set is roughly linear (though a better fit can be obtained by adding a small quadratic coefficient)—nowhere near the $N^4$ dependence of the canonical calculations.

Every notch in accuracy that LNO-CCSD(T) is turned up, from Normal to Tight to vTight, roughly triples CPU times. The same applies to both DLPNO-CCSD(T) and DLPNO-$CCSD(T_1)$. For PNO-LCCSD(T), the timings indicate that having tight thresholds takes ∼1.5 times as long compared to default ones.

We note that for smaller basis sets, LNO Normal, which yields 99.98% of the CCSD(T) correlation energy, comes at a computational cost comparable to DLPNO-$CCSD(T_1)$ TightPNO with TcutPNO = $10^{-6}$ $E_h$; the latter recovers 99% of $E_{corr}$. However, for larger basis sets, LNO becomes more economical.

**Adequacy of def2-nZVP Basis Sets for the Treatment of ($n$ − 1)sp Correlation.** As the present paper was finalized for publication, a reviewer of ref 107 raised an issue that might potentially have some bearing on the present work.

Given the very small gaps between subvalence ($n$ − 1)sp and valence d orbitals in many of these transition metals, Bistoni et al.[108] have argued for the inclusion of these "subvalence" orbitals in "chemical cores" even with the Weigend−Ahlrichs basis sets, and this is indeed the default in the ORCA program system from that group.

As stated above, this practice was adopted also in the present work. While this issue is orthogonal to the relative performance of various localized orbital methods, technically, the def2 basis sets do not include core−valence correlation functions for the transition metals (they do for heavy alkaline and alkaline-earth metals). This raises the question whether this might introduce a significant error in our reference data and whether it would not have been preferable to employ Dunning−Peterson cc-pwCVnZ-PP basis sets instead.[109−113]

A direct comparison between def2-nZVP and cc-pwCVnZ-PP basis set extrapolation would introduce another, less obvious, source of discrepancy: the cc-pwCVnZ-PP basis sets for certain elements use newer-generation relativistic pseudopotentials, which we found can cause discrepancies reaching 1 kcal/mol even at the Hartree−Fock level. However, the differential ($n$ − 1)sp contributions should be directly comparable between the two basis set families. A comparison between def2-QZVPP and cc-pwCVQZ-PP can be found in Table 6, while a similar comparison between the smaller def2-TZVPP and cc-pwCVTZ-PP is shown in Table S8, and for {T,Q} extrapolations, the differences are listed in Table S9 of the SI.

The mean absolute differences between the two sets are 0.29 and 0.27 kcal/mol, respectively, for forward and reverse barriers; the RMS differences are 0.41 and 0.35 kcal/mol. (For {T,Q} extrapolation, these statistics slightly increase to MAD of 0.36 and 0.32 kcal/mol and RMSD of 0.53 and 0.43 kcal/mol, respectively.) The largest single discrepancy with the QZ basis sets is found for reaction **10**, which involves Na in addition to the early second-row transition metal niobium: this reaction is somewhat anomalous, as the (2s, 2p) subvalence orbitals of Na and the (4s, 4p) subvalence orbitals of Nb are nearly degenerate with each other, and the valence−"subvalence" orbital energy gap is a measly 0.2 hartree (!). These numbers are in fact smaller than what was found by Efremenko and Martin[107] for Ru ligand coordination reactions but similar to their statistics for ligand replacement. The broader question of the proper treatment of subvalence correlation in transition metals will be investigated in greater detail, and for broader datasets, in a forthcoming paper.

It is sufficient to say for now that our conclusions on the performance of localized coupled cluster methods are unlikely to be affected by the impact of using def2 basis sets on the description of metal $(n - 1)$sp correlation, as it impacts barrier heights and reaction energies.

Table 6. Effects of Core−Valence Correlation on Forward and Reverse Barriers in LNO-CCSD(T) with Tight Thresholds (kcal/mol)

| | def2-QZVPP | | | | | | cc-pwCVQZ(-PP) | | | | | | | |
|---|---|---|---|---|---|---|---|---|---|---|---|---|---|---|
| | $V_f^‡$, $\Delta E^{\#}_{fwd}$ | | | $V_r^‡$, $\Delta E^{\#}_{rev}$ | | | $V_f^‡$, $\Delta E^{\#}_{fwd}$ | | | $V_r^‡$, $\Delta E^{\#}_{rev}$ | | | $V_f^‡$ | $V_r^‡$ |
| rxn | no $(n-1)$sp | with $(n-1)$sp | $\delta((n-1)$sp$)$ | no $(n-1)$sp | with $(n-1)$sp | $\delta((n-1)$sp$)$ | no $(n-1)$sp | with $(n-1)$sp | $\delta((n-1)$sp$)$ | no $(n-1)$sp | with $(n-1)$sp | $\delta((n-1)$sp$)$ | $\Delta\delta((n-1)$sp$)$[a] | $\Delta\delta((n-1)$sp$)$[a] |
| **1** | 28.76 | 26.17 | −2.59 | 15.89 | 14.28 | −1.61 | 28.71 | 26.47 | −2.24 | 15.91 | 14.82 | −1.09 | 0.35 | 0.52 |
| **2** | 7.42 | 5.81 | −1.61 | 24.35 | 22.31 | −2.04 | 7.36 | 6.29 | −1.08 | 24.23 | 22.37 | −1.86 | 0.54 | 0.18 |
| **3** | 1.22 | 0.95 | −0.27 | 26.69 | 26.98 | 0.29 | 1.19 | 1.00 | −0.19 | 26.61 | 26.26 | −0.36 | 0.08 | −0.65 |
| **4** | 2.07 | 1.45 | −0.62 | 8.21 | 8.29 | 0.08 | 2.00 | 1.69 | −0.30 | 8.19 | 7.82 | −0.37 | 0.31 | −0.45 |
| **5** | 5.31 | 4.86 | −0.46 | 23.01 | 22.57 | −0.45 | 5.24 | 5.01 | −0.23 | 22.79 | 22.62 | −0.17 | 0.23 | 0.28 |
| **6** | 15.64 | 15.72 | 0.08 | 14.94 | 14.83 | −0.11 | 15.42 | 15.44 | 0.02 | 14.92 | 14.82 | −0.10 | −0.05 | 0.02 |
| **7** | 27.69 | 27.68 | −0.01 | 18.85 | 18.79 | −0.05 | 27.67 | 27.64 | −0.03 | 18.85 | 18.78 | −0.07 | −0.02 | −0.01 |
| **8** | 34.97 | 34.20 | −0.77 | 31.71 | 31.50 | −0.21 | $35.01_9$ | $35.01_7$ | $-0.00_2$ | 31.94 | 32.02 | 0.09 | 0.77 | 0.34 |
| **9** | 28.82 | 29.07 | 0.25 | 12.89 | 11.51 | −1.37 | 28.81 | 29.14 | 0.34 | 12.66 | 11.68 | −0.98 | 0.09 | 0.40 |
| **10** | −2.39 | −4.31 | −1.92 | 9.91 | 8.32 | −1.60 | −2.60 | −3.41 | −0.80 | 9.94 | 9.24 | −0.70 | 1.12 | 0.90 |
| **11** | 28.31 | 29.31 | 1.00 | 82.71 | 82.55 | −0.16 | 28.03 | 28.66 | 0.63 | 82.55 | 82.43 | −0.12 | −0.37 | 0.04 |
| **12** | 5.33 | 5.38 | 0.05 | 37.35 | 37.19 | −0.16 | 5.34 | 5.43 | 0.09 | 37.51 | 37.54 | 0.03 | 0.04 | 0.19 |
| **13** | 21.73 | 20.67 | −1.06 | 49.18 | 48.46 | −0.72 | 21.99 | 21.43 | −0.56 | 49.28 | 48.83 | −0.45 | 0.50 | 0.27 |
| **14** | 10.37 | 10.19 | −0.18 | 14.61 | 14.77 | 0.16 | 10.15 | 10.04 | −0.11 | 14.65 | 14.72 | 0.07 | 0.07 | −0.08 |
| **15** | 19.88 | 20.50 | 0.61 | 74.86 | 75.60 | 0.74 | 20.48 | 20.86 | 0.37 | 74.87 | 75.37 | 0.50 | −0.24 | −0.24 |
| **16** | 34.55 | 35.46 | 0.92 | 54.47 | 53.87 | −0.60 | 35.09 | 35.81 | 0.72 | 54.04 | 53.54 | −0.50 | −0.19 | 0.10 |
| **21** | 8.63 | 8.56 | −0.07 | 8.63 | 8.57 | −0.07 | 8.86 | 9.01 | 0.15 | 8.86 | 9.02 | 0.15 | 0.22 | 0.22 |
| **22** | 14.59 | 14.27 | −0.32 | 27.71 | 27.71 | −0.01 | 14.51 | 14.41 | −0.10 | 28.50 | 28.86 | 0.37 | 0.22 | 0.37 |
| **23** | 30.33 | 29.89 | −0.44 | 21.03 | 20.65 | −0.38 | 30.45 | 29.90 | −0.55 | 21.16 | 20.97 | −0.18 | −0.11 | 0.20 |
| **26** | 24.38 | 25.44 | 1.07 | 0.10 | 0.12 | 0.02 | 25.93 | 26.29 | 0.36 | 0.10 | 0.09 | −0.01 | −0.71 | −0.03 |
| **27** | 13.93 | 13.83 | −0.11 | 2.11 | 2.24 | 0.13 | 13.77 | 13.73 | −0.04 | 2.51 | 2.54 | 0.03 | 0.07 | −0.09 |
| **28** | 30.71 | 30.21 | −0.50 | 15.78 | 15.67 | −0.11 | 31.01 | 31.21 | 0.20 | 15.82 | 15.69 | −0.13 | 0.70 | −0.02 |
| **29** | 14.87 | 14.90 | 0.03 | 31.20 | 31.28 | 0.08 | 14.83 | 14.87 | 0.04 | 31.80 | 31.92 | 0.11 | 0.02 | 0.03 |
| **30** | 9.80 | 9.81 | 0.01 | 17.77 | 16.99 | −0.78 | 9.73 | 9.83 | 0.10 | 17.07 | 16.63 | −0.45 | 0.10 | 0.34 |
| **31** | 4.38 | 3.08 | −1.30 | 12.38 | 12.87 | 0.49 | 3.99 | 3.63 | −0.36 | 12.91 | 13.14 | 0.23 | 0.94 | −0.26 |
| **32** | 20.19 | 19.96 | −0.23 | 63.38 | 63.32 | −0.06 | 20.20 | 20.17 | −0.03 | 62.74 | 62.39 | −0.35 | 0.20 | −0.29 |
| **33** | 1.03 | 1.03 | −0.01 | 9.07 | 8.06 | −1.02 | 0.86 | 0.72 | −0.13 | 8.83 | 8.52 | −0.31 | −0.13 | 0.71 |
| **34** | 27.82 | 28.87 | 1.05 | 3.82 | 3.10 | −0.71 | 28.64 | 29.18 | 0.54 | 3.61 | 3.41 | −0.20 | −0.51 | 0.51 |
| **35** | 17.03 | 17.59 | 0.56 | −2.10 | −2.04 | 0.06 | 17.17 | 17.56 | 0.39 | −1.90 | −1.79 | 0.10 | −0.17 | 0.05 |

[a]The $\delta\Delta((n-1)$sp$)$ differences are between LNO-CCSD(T)/cc-pwCVQZ(-PP) and def2-QZVPP for the correlation energies in the forward ($V_f^‡$) and reverse ($V_r^‡$) reactions.

## ■ CONCLUSIONS

We have revisited the MOBH35 (Metal−Organic Barrier Heights) benchmark for realistic organometallic catalytic reactions, using both canonical CCSD(T) and localized orbital approximations to it. For low levels of static correlation, all of DLPNO-CCSD(T), PNO-LCCSD(T), and LNO-CCSD(T) perform well; for moderately strong levels of static correlation, DLPNO-CCSD(T) and ($T_1$) may break down catastrophically, and PNO-LCCSD(T) is vulnerable as well. In contrast, LNO-CCSD(T) converges smoothly to the canonical CCSD(T) answer with increasingly tight convergence settings (Normal, Tight, and vTight). Admittedly, in the kind of case where DLPNO breaks down, the CCSD(T) approximation itself may be questionable, although for this size of a system, we have no computationally tractable way of estimating the impact of connected quadruple excitations. The only two reactions for which our revised MOBH35 reference values differ substantially from the original ones are reaction **9** and to a lesser extent **8**, both involving iron. For the purpose of evaluating DFT methods for MOBH35, it would be best to remove reaction **9** entirely as its severe level of static correlation is just too demanding for a test of any DFT or low-cost wavefunction method. Also, reaction **8** is another electronically challenging case, and we thus advocate its removal as a means of having reference energetics with weak multireference character.

The magnitude of the difference between DLPNO-CCSD(T) and DLPNO-CCSD($T_1$) is a reasonably good predictor for errors in DLPNO-CCSD($T_1$) compared to canonical CCSD-(T); otherwise, monitoring all of $T_1$, $D_1$, max$|t_i^A|$, and $1/(\varepsilon_{LUMO} - \varepsilon_{HOMO})$ should provide adequate warning for potential problems. Our conclusions are not specific to the def2-SVP basis set but are broadly conserved for the larger def2-TZVPP, as they are for the smaller def2-SV(P): the latter may be an economical choice for calibrating against canonical CCSD(T). The diagnostics for static correlation are statistically clustered into groups corresponding to (1) importance of single excitations in the wavefunction; (2a) the small band gap, weakly separated from (2b) correlation entropy; and (3) thermochemical importance of correlation energy, as well as the slope of the DFT reaction energy with respect to the percentage of HF exchange. As a final remark, a variable reduction analysis using the subselect algorithm reveals that the variables that contain much information on the multireference character in MOBH35 are $T_1$, Matito's ratio $I_{ND}/I_{tot}$, and the exchange-based diagnostic $A_{100}$[TPSS].

On a broader note, given a choice between a more rigorous coupled cluster method with a small basis set and a cruder coupled cluster approximation with a more extended basis set, our study indicates that, at least for barrier heights, the former may be the lesser of the two evils. However, both approaches may be combined in a composite scheme along the lines familiar from main-group thermochemistry and noncovalent interactions work.

## ■ ASSOCIATED CONTENT

### Supporting Information

The Supporting Information is available free of charge at https://pubs.acs.org/doi/10.1021/acs.jctc.1c01126.

Box-and-whisker plot for the energy deviations of local CC methods (Figure S1) and MAD statistics for xDSD and ωDSD functionals (Figure S2); data for box plots with quartiles and medians (Tables S1 and S2), frozen core electrons per molecule in MOBH35 (Table S3), total memory requirements (Table S4) and disk usage requirements (Table S5) for local CC single-point calculations, correlation energy recovered from local coupled cluster methods and RMSD values (Table S6), correlation matrix of the MR character diagnostics (Table S7), and effects of core−valence correlation on forward and reverse barriers (Tables S8 and S9) (PDF)

Full raw energies, statistics, and multireference diagnostics (XLSX)

MOBH35 geometries in Cartesian coordinates (TXT)

## ■ AUTHOR INFORMATION

### Corresponding Author

**Jan M.L. Martin** − *Department of Molecular Chemistry and Materials Science, Weizmann Institute of Science, Reḥovot 7610001, Israel;* orcid.org/0000-0002-0005-5074; Email: gershom@weizmann.ac.il

### Author

**Emmanouil Semidalas** − *Department of Molecular Chemistry and Materials Science, Weizmann Institute of Science, Reḥovot 7610001, Israel;* orcid.org/0000-0002-4464-4057

Complete contact information is available at: https://pubs.acs.org/10.1021/acs.jctc.1c01126

### Notes

The authors declare no competing financial interest.

## ■ ACKNOWLEDGMENTS

This research was supported by the Israel Science Foundation (grant 1969/20), the Minerva Foundation (grant 2020/05), and the Helen and Martin Kimmel Center for Molecular Design (Weizmann Institute of Science). The work of E.S. on this scientific paper was supported by the Onassis Foundation—Scholarship ID: F ZP 052-1/2019-2020. We thank Drs. Mark A. Iron and Irena Efremenko for helpful discussions and Dr. Andreas Hansen and Prof. Frank Neese for clarifications concerning the VeryTightPNO criteria.

## ■ DEDICATION

J.M.L.M. would like to dedicate this paper to the memory of his longtime colleague Shimon Vega (1943−2021), leading light of Israel's NMR community, educator in heart and soul, and mainstay of the Faculty of Chemistry.

## ■ REFERENCES

(1) Gunanathan, C.; Milstein, D. Bond Activation and Catalysis by Ruthenium Pincer Complexes. *Chem. Rev.* **2014**, *114*, 12024−12087.

(2) Zhan, C.; Lian, C.; Zhang, Y.; Thompson, M. W.; Xie, Y.; Wu, J.; Kent, P. R. C.; Cummings, P. T.; Jiang, D.; Wesolowski, D. J. Computational Insights into Materials and Interfaces for Capacitive Energy Storage. *Adv. Sci.* **2017**, *4*, 1700059.

(3) Yang, Z.; Mehmood, R.; Wang, M.; Qi, H. W.; Steeves, A. H.; Kulik, H. J. Revealing Quantum Mechanical Effects in Enzyme Catalysis with Large-Scale Electronic Structure Simulation. *React. Chem. Eng.* **2019**, *4*, 298−315.

(4) Crawley, M. L.; Trost, B. M. Applications of Transition Metal Catalysis *in Drug Discovery and Development*; Crawley, M. L., Trost, B. M., Eds.; John Wiley & Sons, Inc.: Hoboken, NJ, USA, 2012.
(5) Vogiatzis, K. D.; Polynski, M. V.; Kirkland, J. K.; Townsend, J.; Hashemi, A.; Liu, C.; Pidko, E. A. Computational Approach to Molecular Catalysis by 3d Transition Metals: Challenges and Opportunities. *Chem. Rev.* **2019**, *119*, 2453−2523.
(6) Raghavachari, K.; Trucks, G. W.; Pople, J. A.; Head-Gordon, M. A Fifth-Order Perturbation Comparison of Electron Correlation Theories. *Chem. Phys. Lett.* **1989**, *157*, 479−483.
(7) Watts, J. D.; Gauss, J.; Bartlett, R. J. Coupled-cluster Methods with Noniterative Triple Excitations for Restricted Open-shell Hartree−Fock and Other General Single Determinant Reference Functions. Energies and Analytical Gradients. *J. Chem. Phys.* **1993**, *98*, 8718−8733.
(8) Bak, K. L.; Jørgensen, P.; Olsen, J.; Helgaker, T.; Gauss, J. Coupled-Cluster Singles, Doubles and Triples (CCSDT) Calculations of Atomization Energies. *Chem. Phys. Lett.* **2000**, *317*, 116−122.
(9) Ruden, T. A.; Helgaker, T.; Jørgensen, P.; Olsen, J. Coupled-Cluster Connected-Quadruples Corrections to Atomization Energies. *Chem. Phys. Lett.* **2003**, *371*, 62−67.
(10) Ruden, T. A.; Helgaker, T.; Jørgensen, P.; Olsen, J. Coupled-Cluster Connected Quadruples and Quintuples Corrections to the Harmonic Vibrational Frequencies and Equilibrium Bond Distances of HF, $N_2$, $F_2$, and CO. *J. Chem. Phys.* **2004**, *121*, 5874−5884.
(11) Daniel Boese, A.; Oren, M.; Atasoylu, O.; Martin, J. M. L.; Kállay, M.; Gauss, J. W3 Theory: Robust Computational Thermochemistry in the KJ/Mol Accuracy Range. *J. Chem. Phys.* **2004**, 3405.
(12) Tajti, A.; Szalay, P. G.; Császár, A. G.; Kállay, M.; Gauss, J.; Valeev, E. F.; Flowers, B. A.; Vázquez, J.; Stanton, J. F. HEAT: High Accuracy Extrapolated Ab Initio Thermochemistry. *J. Chem. Phys.* **2004**, *121*, 11599−11613.
(13) Urban, M.; Noga, J.; Cole, S. J.; Bartlett, R. J. Towards a Full CCSDT Model for Electron Correlation. *J. Chem. Phys.* **1985**, *83*, 4041−4046.
(14) Bomble, Y. J.; Stanton, J. F.; Kállay, M.; Gauss, J. Coupled-Cluster Methods Including Noniterative Corrections for Quadruple Excitations. *J. Chem. Phys.* **2005**, *123*, 054101.
(15) Oliphant, N.; Adamowicz, L. Coupled-Cluster Method Truncated at Quadruples. *J. Chem. Phys.* **1991**, *95*, 6645−6651.
(16) Kucharski, S. A.; Bartlett, R. J. The Coupled-Cluster Single, Double, Triple, and Quadruple Excitation Method. *J. Chem. Phys.* **1992**, *97*, 4282−4288.
(17) Stanton, J. F. Why CCSD(T) Works: A Different Perspective. *Chem. Phys. Lett.* **1997**, *281*, 130−134.
(18) Riplinger, C.; Neese, F. An Efficient and near Linear Scaling Pair Natural Orbital Based Local Coupled Cluster Method. *J. Chem. Phys.* **2013**, *138*, 034106.
(19) Riplinger, C.; Sandhoefer, B.; Hansen, A.; Neese, F. Natural Triple Excitations in Local Coupled Cluster Calculations with Pair Natural Orbitals. *J. Chem. Phys.* **2013**, *139*, 134101.
(20) Ma, Q.; Werner, H.-J. J. Explicitly Correlated Local Coupled-Cluster Methods Using Pair Natural Orbitals. *Wiley Interdiscip. Rev.: Comput. Mol. Sci.* **2018**, *8*, e1371.
(21) Rolik, Z.; Szegedy, L.; Ladjánszki, I.; Ladóczki, B.; Kállay, M. An Efficient Linear-Scaling CCSD(T) Method Based on Local Natural Orbitals. *J. Chem. Phys.* **2013**, *139*, 094105.
(22) Schmitz, G.; Hättig, C.; Tew, D. P. Explicitly Correlated PNO-MP2 and PNO-CCSD and Their Application to the S66 Set and Large Molecular Systems. *Phys. Chem. Chem. Phys.* **2014**, *16*, 22167−22178.
(23) Schmitz, G.; Hättig, C. Perturbative Triples Correction for Local Pair Natural Orbital Based Explicitly Correlated CCSD(F12*) Using Laplace Ransformation Techniques. *J. Chem. Phys.* **2016**, *145*, 234107.
(24) Semidalas, E.; Martin, J. M. L. Canonical and DLPNO-Based G4(MP2)XK-Inspired Composite Wave Function Methods Parametrized against Large and Chemically Diverse Training Sets: Are They More Accurate and/or Robust than Double-Hybrid DFT? *J. Chem. Theory Comput.* **2020**, *16*, 4238−4255.
(25) Semidalas, E.; Martin, J. M. L. Canonical and DLPNO-Based Composite Wavefunction Methods Parametrized against Large and Chemically Diverse Training Sets. 2: Correlation-Consistent Basis Sets, Core−Valence Correlation, and F12 Alternatives. *J. Chem. Theory Comput.* **2020**, *16*, 7507−7524.
(26) Goerigk, L.; Hansen, A.; Bauer, C.; Ehrlich, S.; Najibi, A.; Grimme, S. A Look at the Density Functional Theory Zoo with the Advanced GMTKN55 Database for General Main Group Thermochemistry, Kinetics and Noncovalent Interactions. *Phys. Chem. Chem. Phys.* **2017**, *19*, 32184−32215.
(27) Aoto, Y. A.; de Lima Batista, A. P.; Köhn, A.; de Oliveira-Filho, A. G. S. How To Arrive at Accurate Benchmark Values for Transition Metal Compounds: Computation or Experiment? *J. Chem. Theory Comput.* **2017**, *13*, 5291−5316.
(28) Xu, X.; Zhang, W.; Tang, M.; Truhlar, D. G. Do Practical Standard Coupled Cluster Calculations Agree Better than Kohn-Sham Calculations with Currently Available Functionals When Compared to the Best Available Experimental Data for Dissociation Energies of Bonds to 3d Transition Metals? *J. Chem. Theory Comput.* **2015**, *11*, 2036−2052.
(29) Chan, B. The CUAGAU Set of Coupled-Cluster Reference Data for Small Copper, Silver, and Gold Compounds and Assessment of DFT Methods. *J. Phys. Chem. A* **2019**, *123*, 5781−5788.
(30) Chan, B. Assessment and Development of DFT with the Expanded CUAGAU-2 Set of Group-11 Cluster Systems. *Int. J. Quantum Chem.* **2021**, *121*, e26453.
(31) Hait, D.; Tubman, N. M.; Levine, D. S.; Whaley, K. B.; Head-Gordon, M. What Levels of Coupled Cluster Theory Are Appropriate for Transition Metal Systems? A Study Using Near-Exact Quantum Chemical Values for 3d Transition Metal Binary Compounds. *J. Chem. Theory Comput.* **2019**, *15*, 5370−5385.
(32) Dohm, S.; Hansen, A.; Steinmetz, M.; Grimme, S.; Checinski, M. P. Comprehensive Thermochemical Benchmark Set of Realistic Closed-Shell Metal Organic Reactions. *J. Chem. Theory Comput.* **2018**, *14*, 2596−2608.
(33) Iron, M. A.; Janes, T. Evaluating Transition Metal Barrier Heights with the Latest Density Functional Theory Exchange−Correlation Functionals: The MOBH35 Benchmark Database. *J. Phys. Chem. A* **2019**, *123*, 3761−3781.
(34) Iron, M. A.; Janes, T. Correction to "Evaluating Transition Metal Barrier Heights with the Latest Density Functional Theory Exchange−Correlation Functionals: The MOBH35 Benchmark Database". *J. Phys. Chem. A* **2019**, *123*, 6379−6380.
(35) Liakos, D. G.; Sparta, M.; Kesharwani, M. K.; Martin, J. M. L.; Neese, F. Exploring the Accuracy Limits of Local Pair Natural Orbital Coupled-Cluster Theory. *J. Chem. Theory Comput.* **2015**, *11*, 1525−1539.
(36) Guo, Y.; Riplinger, C.; Becker, U.; Liakos, D. G.; Minenkov, Y.; Cavallo, L.; Neese, F. Communication: An Improved Linear Scaling Perturbative Triples Correction for the Domain Based Local Pair-Natural Orbital Based Singles and Doubles Coupled Cluster Method [DLPNO-CCSD(T)]. *J. Chem. Phys.* **2018**, *148*, 011101.
(37) Maurer, L. R.; Bursch, M.; Grimme, S.; Hansen, A. Assessing Density Functional Theory for Chemically Relevant Open-Shell Transition Metal Reactions. *J. Chem. Theory Comput.* **2021**, *17*, 6134−6151.
(38) Sylvetsky, N.; Banerjee, A.; Alonso, M.; Martin, J. M. L. Performance of Localized Coupled Cluster Methods in a Moderately Strong Correlation Regime: Hückel−Möbius Interconversions in Expanded Porphyrins. *J. Chem. Theory Comput.* **2020**, *16*, 3641−3653.
(39) Rolik, Z.; Kállay, M. A General-Order Local Coupled-Cluster Method Based on the Cluster-in-Molecule Approach. *J. Chem. Phys.* **2011**, *135*, 104111.
(40) Kállay, M. Linear-Scaling Implementation of the Direct Random-Phase Approximation. *J. Chem. Phys.* **2015**, *142*, 204105.
(41) Nagy, P. R.; Kállay, M. Optimization of the Linear-Scaling Local Natural Orbital CCSD(T) Method: Redundancy-Free Triples Correction Using Laplace Transform. *J. Chem. Phys.* **2017**, *146*, 214106.
(42) Nagy, P. R.; Samu, G.; Kállay, M. Optimization of the Linear-Scaling Local Natural Orbital CCSD(T) Method: Improved Algorithm

and Benchmark Applications. *J. Chem. Theory Comput.* **2018**, *14*, 4193−4215.
(43) Schwilk, M.; Usvyat, D.; Werner, H.-J. Communication: Improved Pair Approximations in Local Coupled-Cluster Methods. *J. Chem. Phys.* **2015**, *142*, 121102.
(44) Ma, Q.; Werner, H.-J. Scalable Electron Correlation Methods. 5. Parallel Perturbative Triples Correction for Explicitly Correlated Local Coupled Cluster with Pair Natural Orbitals. *J. Chem. Theory Comput.* **2018**, *14*, 198−215.
(45) Dohm, S.; Bursch, M.; Hansen, A.; Grimme, S. Semiautomated Transition State Localization for Organometallic Complexes with Semiempirical Quantum Chemical Methods. *J. Chem. Theory Comput.* **2020**, *16*, 2002−2012.
(46) Weigend, F.; Ahlrichs, R. Balanced Basis Sets of Split Valence, Triple Zeta Valence and Quadruple Zeta Valence Quality for H to Rn: Design and Assessment of Accuracy. *Phys. Chem. Chem. Phys.* **2005**, *7*, 3297−3305.
(47) Dolg, M.; Cao, X. Relativistic Pseudopotentials: Their Development and Scope of Applications. *Chem. Rev.* **2012**, *112*, 403−480.
(48) Weigend, F. Hartree−Fock Exchange Fitting Basis Sets for H to Rn. *J. Comput. Chem.* **2008**, *29*, 167−175.
(49) Hellweg, A.; Hättig, C.; Höfener, S.; Klopper, W. Optimized Accurate Auxiliary Basis Sets for RI-MP2 and RI-CC2 Calculations for the Atoms Rb to Rn. *Theor. Chem. Acc.* **2007**, *117*, 587−597.
(50) Weigend, F.; Häser, M.; Patzelt, H.; Ahlrichs, R. RI-MP2: Optimized Auxiliary Basis Sets and Demonstration of Efficiency. *Chem. Phys. Lett.* **1998**, *294*, 143−152.
(51) Werner, H.-J.; Knowles, P. J.; Manby, F. R.; Black, J. A.; Doll, K.; Heßelmann, A.; Kats, D.; Köhn, A.; Korona, T.; Kreplin, D. A.; Ma, Q.; Miller, T. F., III; Mitrushchenkov, A.; Peterson, K. A.; Polyak, I.; Rauhut, G.; Sibaev, M. The Molpro Quantum Chemistry Package. *J. Chem. Phys.* **2020**, *152*, 144107.
(52) Deprince, A. E., III; Sherrill, C. D. Accuracy and Efficiency of Coupled-Cluster Theory Using Density Fitting/Cholesky Decomposition, Frozen Natural Orbitals, and a $t_1$-Transformed Hamiltonian. *J. Chem. Theory Comput.* **2013**, *9*, 2687−2696.
(53) Smith, D. G. A.; Burns, L. A.; Simmonett, A. C.; Parrish, R. M.; Schieber, M. C.; Galvelis, R.; Kraus, P.; Kruse, H.; Di Remigio, R.; Alenaizan, A.; James, A. M.; Lehtola, S.; Misiewicz, J. P.; Scheurer, M.; Shaw, R. A.; Schriber, J. B.; Xie, Y.; Glick, Z. L.; Sirianni, D. A.; O'Brien, J. S.; Waldrop, J. M.; Kumar, A.; Hohenstein, E. G.; Pritchard, B. P.; Brooks, B. R.; Schaefer, H. F.; Sokolov, A. Y.; Patkowski, K.; DePrince, A. E.; Bozkaya, U.; King, R. A.; Evangelista, F. A.; Turney, J. M.; Crawford, T. D.; Sherrill, C. D. PSI4 1.4: Open-Source Software for High-Throughput Quantum Chemistry. *J. Chem. Phys.* **2020**, *152*, 184108.
(54) Gyevi-Nagy, L.; Kállay, M.; Nagy, P. R. Integral-Direct and Parallel Implementation of the CCSD(T) Method: Algorithmic Developments and Large-Scale Applications. *J. Chem. Theory Comput.* **2020**, *16*, 366−384.
(55) Kállay, M.; Nagy, P. R.; Mester, D.; Rolik, Z.; Samu, G.; Csontos, J.; Csóka, J.; Szabó, P. B.; Gyevi-Nagy, L.; Hégely, B.; Ladjánszki, I.; Szegedy, L.; Ladóczki, B.; Petrov, K.; Farkas, M.; Mezei, P. D.; Ganyecz, Á. The MRCC Program System: Accurate Quantum Chemistry from Water to Proteins. *J. Chem. Phys.* **2020**, *152*, 074107.
(56) Frisch, M. J.; Trucks, G. W.; Schlegel, H. B.; Scuseria, G. E.; Robb, M. A.; Cheeseman, J. R.; Scalmani, G.; Barone, V.; Petersson, G. A.; Nakatsuji, H.; Li, X.; Caricato, M.; Marenich, A. V.; Bloino, J.; Janesko, B. G.; Gomperts, R.; Mennucci, B.; Hratchian, H. P.; Ortiz, J. V.; Izmaylov, A. F.; Sonnenberg, J. L.; Williams-Young, D.; Ding, F.; Lipparini, F.; Egidi, F.; Goings, J.; Peng, B.; Petrone, A.; Henderson, T.; Ranasinghe, D.; Zakrzewski, V. G.; Gao, J.; Rega, N.; Zheng, G.; Liang, W.; Hada, M.; Ehara, M.; Toyota, K.; Fukuda, R.; Hasegawa, J.; Ishida, M.; Nakajima, T.; Honda, Y.; Kitao, O.; Nakai, H.; Vreven, T.; Throssell, K.; Montgomery, J. A. J.; Peralta, J. E.; Ogliaro, F.; Bearpark, M. J.; Heyd, J. J.; Brothers, E. N.; Kudin, K. N.; Staroverov, V. N.; Keith, T. A.; Kobayashi, R.; Normand, J.; Raghavachari, K.; Rendell, A. P.; Burant, J. C.; Iyengar, S. S.; Tomasi, J. C.; Cossi, M.; Millam, J. M.; Klene, M.; Adamo, C.; Cammi, R.; Ochterski, J. W.; Martin, R. L.; Morokuma, K.; Farkas, O.; Foresman, J. B.; Fox, D. J. *Gaussian 16*, Rev. C.01. Gaussian, Inc., Wallingford, CT. 2016.
(57) Schwilk, M.; Ma, Q.; Köppl, C.; Werner, H.-J. Scalable Electron Correlation Methods. 3. Efficient and Accurate Parallel Local Coupled Cluster with Pair Natural Orbitals (PNO-LCCSD). *J. Chem. Theory Comput.* **2017**, *13*, 3650−3675.
(58) Ma, Q.; Schwilk, M.; Köppl, C.; Werner, H.-J. Scalable Electron Correlation Methods. 4. Parallel Explicitly Correlated Local Coupled Cluster with Pair Natural Orbitals (PNO-LCCSD-F12). *J. Chem. Theory Comput.* **2017**, *13*, 4871−4896.
(59) Nagy, P. R.; Kállay, M. Approaching the Basis Set Limit of CCSD(T) Energies for Large Molecules with Local Natural Orbital Coupled-Cluster Methods. *J. Chem. Theory Comput.* **2019**, *15*, 5275−5298.
(60) Saitow, M.; Becker, U.; Riplinger, C.; Valeev, E. F.; Neese, F. A New Near-Linear Scaling, Efficient and Accurate, Open-Shell Domain-Based Local Pair Natural Orbital Coupled Cluster Singles and Doubles Theory. *J. Chem. Phys.* **2017**, *146*, 164105.
(61) Neese, F.; Wennmohs, F.; Becker, U.; Riplinger, C. The ORCA Quantum Chemistry Program Package. *J. Chem. Phys.* **2020**, *152*, 224108.
(62) Pavošević, F.; Peng, C.; Pinski, P.; Riplinger, C.; Neese, F.; Valeev, E. F. SparseMaps—A Systematic Infrastructure for Reduced Scaling Electronic Structure Methods. V. Linear Scaling Explicitly Correlated Coupled-Cluster Method with Pair Natural Orbitals. *J. Chem. Phys.* **2017**, *146*, 174108.
(63) Kossmann, S.; Neese, F. Efficient Structure Optimization with Second-Order Many-Body Perturbation Theory: The RIJCOSX-MP2 Method. *J. Chem. Theory Comput.* **2010**, *6*, 2325−2338.
(64) Neese, F.; Wennmohs, F.; Hansen, A.; Becker, U. Efficient, Approximate and Parallel Hartree-Fock and Hybrid DFT Calculations. A "chain-of-Spheres" Algorithm for the Hartree-Fock Exchange. *Chem. Phys.* **2009**, *356*, 98−109.
(65) Neese, F.; Valeev, E. F. Revisiting the Atomic Natural Orbital Approach for Basis Sets: Robust Systematic Basis Sets for Explicitly Correlated and Conventional Correlated Ab Initio Methods? *J. Chem. Theory Comput.* **2011**, *7*, 33−43.
(66) Schwartz, C. Ground State of the Helium Atom. *Phys. Rev.* **1962**, *128*, 1146−1148.
(67) Schwartz, C. Importance of Angular Correlations between Atomic Electrons. *Phys. Rev.* **1962**, *126*, 1015−1019.
(68) Hill, R. N. Rates of Convergence and Error Estimation Formulas for the Rayleigh−Ritz Variational Method. *J. Chem. Phys.* **1985**, *83*, 1173−1196.
(69) Kutzelnigg, W.; Morgan, J. D., III Rates of Convergence of the Partial-wave Expansions of Atomic Correlation Energies. *J. Chem. Phys.* **1992**, *96*, 4484−4508.
(70) Martin, J. M. L. Ab Initio Total Atomization Energies of Small Molecules — towards the Basis Set Limit. *Chem. Phys. Lett.* **1996**, *259*, 669−678.
(71) Halkier, A.; Helgaker, T.; Jørgensen, P.; Klopper, W.; Koch, H.; Olsen, J.; Wilson, A. K. Basis-Set Convergence in Correlated Calculations on Ne, $N_2$, and $H_2O$. *Chem. Phys. Lett.* **1998**, *286*, 243−252.
(72) Karton, A.; Martin, J. M. L. Comment on: "Estimating the Hartree−Fock Limit from Finite Basis Set Calculations" [Jensen F (2005) Theor Chem Acc 113:267]. *Theor. Chem. Acc.* **2006**, *115*, 330−333.
(73) Klopper, W.; Kutzelnigg, W. Gaussian Basis Sets and the Nuclear Cusp Problem. *J. Mol. Struct.: THEOCHEM* **1986**, *135*, 339−356.
(74) McKemmish, L. K.; Gill, P. M. W. Gaussian Expansions of Orbitals. *J. Chem. Theory Comput.* **2012**, *8*, 4891−4898.
(75) Kutzelnigg, W. Theory of the Expansion of Wave Functions in a Gaussian Basis. *Int. J. Quantum Chem.* **1994**, *51*, 447−463.
(76) Schwenke, D. W. The Extrapolation of One-Electron Basis Sets in Electronic Structure Calculations: How It Should Work and How It Can Be Made to Work. *J. Chem. Phys.* **2005**, *122*, 014107.

(77) Martin, J. M. L. A Simple "Range Extender" for Basis Set Extrapolation Methods for MP2 and Coupled Cluster Correlation Energies. *AIP Conf. Proc.* **2018**, *2040*, 020008.

(78) Lee, T. J.; Taylor, P. R. A Diagnostic for Determining the Quality of Single-Reference Electron Correlation Methods. *Int. J. Quantum Chem.* **1989**, *36*, 199−207.

(79) Janssen, C. L.; Nielsen, I. M. B. New Diagnostics for Coupled-Cluster and Møller−Plesset Perturbation Theory. *Chem. Phys. Lett.* **1998**, *290*, 423−430.

(80) Leininger, M. L.; Nielsen, I. M. B.; Crawford, T. D.; Janssen, C. L. A New Diagnostic for Open-Shell Coupled-Cluster Theory. *Chem. Phys. Lett.* **2000**, *328*, 431−436.

(81) Lee, T. J. Comparison of the $T_1$ and $D_1$ Diagnostics for Electronic Structure Theory: A New Definition for the Open-Shell $D_1$ Diagnostic. *Chem. Phys. Lett.* **2003**, *372*, 362−367.

(82) Fogueri, U. R.; Kozuch, S.; Karton, A.; Martin, J. M. L. A Simple DFT-Based Diagnostic for Nondynamical Correlation. *Theor. Chem. Acc.* **2012**, *132*, 1−9.

(83) Hollett, J. W.; Gill, P. M. W. The Two Faces of Static Correlation. *J. Chem. Phys.* **2011**, *134*, 114111.

(84) Coulson, C. A.; Fischer, I. XXXIV. Notes on the Molecular Orbital Treatment of the Hydrogen Molecule. *London, Edinburgh, Dublin Philos. Mag. J. Sci.* **1949**, *40*, 386−393.

(85) Mermin, N. D. Thermal Properties of the Inhomogeneous Electron Gas. *Phys. Rev.* **1965**, *137*, A1441−A1443.

(86) Rabuck, A. D.; Scuseria, G. E. Improving Self-Consistent Field Convergence by Varying Occupation Numbers. *J. Chem. Phys.* **1999**, *110*, 695−700.

(87) Karton, A.; Rabinovich, E.; Martin, J. M. L.; Ruscic, B. W4 Theory for Computational Thermochemistry: In Pursuit of Confident Sub-KJ/Mol Predictions. *J. Chem. Phys.* **2006**, *125*, 144108.

(88) Karton, A.; Taylor, P. R.; Martin, J. M. L. Basis Set Convergence of Post-CCSD Contributions to Molecular Atomization Energies. *J. Chem. Phys.* **2007**, *127*, 064104.

(89) Karton, A.; Daon, S.; Martin, J. M. L. W4-11: A High-Confidence Benchmark Dataset for Computational Thermochemistry Derived from First-Principles W4 Data. *Chem. Phys. Lett.* **2011**, *510*, 165−178.

(90) Tishchenko, O.; Zheng, J.; Truhlar, D. G. Multireference Model Chemistries for Thermochemical Kinetics. *J. Chem. Theory Comput.* **2008**, *4*, 1208−1219.

(91) Ramos-Cordoba, E.; Salvador, P.; Matito, E. Separation of Dynamic and Nondynamic Correlation. *Phys. Chem. Chem. Phys.* **2016**, *18*, 24015−24023.

(92) Ramos-Cordoba, E.; Matito, E. Local Descriptors of Dynamic and Nondynamic Correlation. *J. Chem. Theory Comput.* **2017**, *13*, 2705−2711.

(93) Kesharwani, M. K.; Sylvetsky, N.; Köhn, A.; Tew, D. P.; Martin, J. M. L. Do CCSD and Approximate CCSD-F12 Variants Converge to the Same Basis Set Limits? The Case of Atomization Energies. *J. Chem. Phys.* **2018**, *149*, 154109.

(94) Xu, X.; Sitkiewicz, S.; Ramos-cordoba, E.; Lopez, X.; Matito, E. Analysis of Dynamic and Nondynamic Correlation Diagnostics. In *Math/Chem/Comp 2021 - 32nd MC2 Conference,* Inter University Centre Dubrovnik, *7−11 June, 2021*; 2021; p 21.

(95) Grimme, S.; Hansen, A. A Practicable Real-Space Measure and Visualization of Static Electron-Correlation Effects. *Angew. Chem., Int. Ed.* **2015**, *54*, 12308−12313.

(96) Martin, J. M. L.; Santra, G.; Semidalas, E. An Exchange-Based Diagnostic for Static Correlation. *AIP Conf. Proc.* **2021**, *in press*, arXiv:2111.01879.

(97) Handy, N. C.; Cohen, A. J. Left-Right Correlation Energy. *Mol. Phys.* **2001**, *99*, 403−412.

(98) Martin, J. M. L.; de Oliveira, G. Towards Standard Methods for Benchmark Quality Ab Initio Thermochemistry—W1 and W2 Theory. *J. Chem. Phys.* **1999**, *111*, 1843−1856.

(99) Hoaglin, D. C.; Mosteller, F.; Tukey, J. W. *Fundamentals of Exploratory Analysis of Variance*; Wiley Series in Probability and Statistics; Wiley, 1991.

(100) Santra, G.; Cho, M.; Martin, J. M. L. Exploring Avenues beyond Revised DSD Functionals: I. Range Separation, with x DSD as a Special Case. *J. Phys. Chem. A* **2021**, *125*, 4614−4627.

(101) Santra, G.; Semidalas, E.; Martin, J. M. L. Exploring Avenues Beyond Revised DSD Functionals: II. Random-Phase Approximation and Scaled MP3 Corrections. *J. Phys. Chem. A* **2021**, *125*, 4628−4638.

(102) Altun, A.; Neese, F.; Bistoni, G. Extrapolation to the Limit of a Complete Pair Natural Orbital Space in Local Coupled-Cluster Calculations. *J. Chem. Theory Comput.* **2020**, *16*, 6142−6149.

(103) Semidalas, E.; Santra, G.; Mehta, N.; Martin, J. M. L. The S66 Noncovalent Interaction Benchmark Re-Examined: Composite Localized Coupled Cluster Approaches. *AIP Conf. Proc.* **2021**, *in press*, arXiv:2111.01882.

(104) Karton, A.; Sylvetsky, N.; Martin, J. M. L. W4-17: A Diverse and High-Confidence Dataset of Atomization Energies for Benchmarking High-Level Electronic Structure Methods. *J. Comput. Chem.* **2017**, *38*, 2063−2075.

(105) Cadima, J.; Cerdeira, J. O.; Minhoto, M. Computational Aspects of Algorithms for Variable Selection in the Context of Principal Components. *Comput. Stat. Data Anal.* **2004**, *47*, 225−236.

(106) R Core Team. *R: A Language and Environment for Statistical Computing*. Vienna, Austria 2021.

(107) Efremenko, I.; Martin, J. M. L. Coupled Cluster Benchmark of New DFT and Local Correlation Methods: Mechanisms of Hydro-arylation and Oxidative Coupling Catalyzed by Ru(II, III) Chloride Carbonyls. *J. Phys. Chem. A* **2021**, *125*, 8987−8999.

(108) Bistoni, G.; Riplinger, C.; Minenkov, Y.; Cavallo, L.; Auer, A. A.; Neese, F. Treating Subvalence Correlation Effects in Domain Based Pair Natural Orbital Coupled Cluster Calculations: An Out-of-the-Box Approach. *J. Chem. Theory Comput.* **2017**, *13*, 3220−3227.

(109) Balabanov, N. B.; Peterson, K. A. Systematically Convergent Basis Sets for Transition Metals. I. All-Electron Correlation Consistent Basis Sets for the 3d Elements Sc−Zn. *J. Chem. Phys.* **2005**, *123*, 064107.

(110) Figgen, D.; Peterson, K. A.; Dolg, M.; Stoll, H. Energy-Consistent Pseudopotentials and Correlation Consistent Basis Sets for the 5d Elements Hf−Pt. *J. Chem. Phys.* **2009**, *130*, 164108.

(111) Peterson, K. A.; Puzzarini, C. Systematically Convergent Basis Sets for Transition Metals. II. Pseudopotential-Based Correlation Consistent Basis Sets for the Group 11 (Cu, Ag, Au) and 12 (Zn, Cd, Hg) Elements. *Theor. Chem. Acc.* **2005**, *114*, 283−296.

(112) Figgen, D.; Rauhut, G.; Dolg, M.; Stoll, H. Energy-Consistent Pseudopotentials for Group 11 and 12 Atoms: Adjustment to Multi-Configuration Dirac−Hartree−Fock Data. *Chem. Phys.* **2005**, *311*, 227−244.

(113) Peterson, K. A.; Figgen, D.; Dolg, M.; Stoll, H. Energy-Consistent Relativistic Pseudopotentials and Correlation Consistent Basis Sets for the 4d Elements Y−Pd. *J. Chem. Phys.* **2007**, *126*, 124101.